\documentclass[letterpaper]{article} 
\usepackage{aaai24}  
\usepackage{times}  
\usepackage{helvet}  
\usepackage{courier}  
\usepackage[hyphens]{url}  
\usepackage{graphicx} 
\usepackage{natbib}  
\usepackage{caption} 
\usepackage{algorithm}
\usepackage{algorithmic}
\usepackage{multirow}
\usepackage{booktabs}
\usepackage{amsmath}
\usepackage{amssymb}
\usepackage{newfloat}
\usepackage{listings}
\DeclareCaptionStyle{ruled}{labelfont=normalfont,labelsep=colon,strut=off} 
\floatstyle{ruled}
\newfloat{listing}{tb}{lst}{}
\floatname{listing}{Listing}
\title{Does Few-shot Learning Suffer from Backdoor Attacks?}
\author{
    Xinwei Liu\textsuperscript{\rm 1,2} , Xiaojun Jia\textsuperscript{\rm 3}\thanks{Correspondence to: Xiaojun Jia and Xiaochun Cao.}, Jindong Gu\textsuperscript{\rm 4}, Yuan Xun\textsuperscript{\rm 1,2}\\
    Siyuan Liang\textsuperscript{\rm 5}, Xiaochun Cao\textsuperscript{\rm 6*}
}
\affiliations{
    \textsuperscript{\rm 1}SKLOIS, Institute of Information Engineering, CAS, Beijing, China\\
    \textsuperscript{\rm 2}School of Cyber Security, University of Chinese Academy of Sciences, Beijing, China\\
    \textsuperscript{\rm 3}Nanyang Technological University, Singapore\\
    \textsuperscript{\rm 4}University of Oxford, UK\\
    \textsuperscript{\rm 5}School of Computing, National University of Singapore, Singapore\\
    \textsuperscript{\rm 6}School of Cyber Science and Technology, Shenzhen Campus, Sun Yat-sen University, Shenzhen, China
    \{liuxinwei,xunyuan\}@iie.ac.cn, jiaxiaojunqaq@gmail.com, 
    jindong.gu@outlook.com,\\
    pandaliang521@gmail.com, caoxiaochun@mail.sysu.edu.cn

}

\usepackage{bibentry}

\begin{document}

\maketitle

\begin{abstract}
The field of few-shot learning (FSL) has shown promising results in scenarios where training data is limited, but its vulnerability to backdoor attacks remains largely unexplored. We first explore this topic by first evaluating the performance of the existing backdoor attack methods on few-shot learning scenarios. Unlike in standard supervised learning, existing backdoor attack methods failed to perform an effective attack in FSL due to two main issues. Firstly, the model tends to overfit to either benign features or trigger features, causing a tough trade-off between attack success rate and benign accuracy. Secondly, due to the small number of training samples, the dirty label or visible trigger in the support set can be easily detected by victims, which reduces the stealthiness of attacks. It seemed that FSL could survive from backdoor attacks.  However, in this paper, we propose the Few-shot Learning Backdoor Attack (FLBA) to show that FSL can still be vulnerable to backdoor attacks. Specifically, we first generate a trigger to maximize the gap between poisoned and benign features. It enables the model to learn both benign and trigger features, which solves the problem of overfitting. To make it more stealthy, we hide the trigger by optimizing two types of imperceptible perturbation, namely attractive and repulsive perturbation, instead of attaching the trigger directly. Once we obtain the perturbations, we can poison all samples in the benign support set into a hidden poisoned support set and fine-tune the model on it. Our method demonstrates a high Attack Success Rate (ASR) in FSL tasks with different few-shot learning paradigms while preserving clean accuracy and maintaining stealthiness. This study reveals that few-shot learning still suffers from backdoor attacks, and its security should be given attention.

\end{abstract}


\section{Introduction}
Deep learning has demonstrated remarkable success in a variety of computer vision applications, particularly in the area of image classification~\cite{he2016deep,gu2018recent}. However, this success is contingent upon access to a significant amount of training data. In many real-world scenarios, only a few labeled samples are available for new unseen classes, such as rare species or medical diseases.  Learning a classifier when only a few training samples are available for each class is well known as few-shot learning (FSL) in the literature~\cite{vinyals2016matching,finn2017model}. The FSL approach involves using a large auxiliary set of labeled data from disjoint classes to acquire transferable knowledge or representations that can help in the few-shot tasks. Recently, the security implications of FSL have been brought to the forefront of the community~\cite{li2022few,guan2022few}, such as the challenge of training a robust few-shot model against adversarial attacks~\cite{li2019defensive,jia2020adv,huang2021advfilter,huang2023ala}.

\begin{figure}
\centering
    \includegraphics[width=\linewidth]{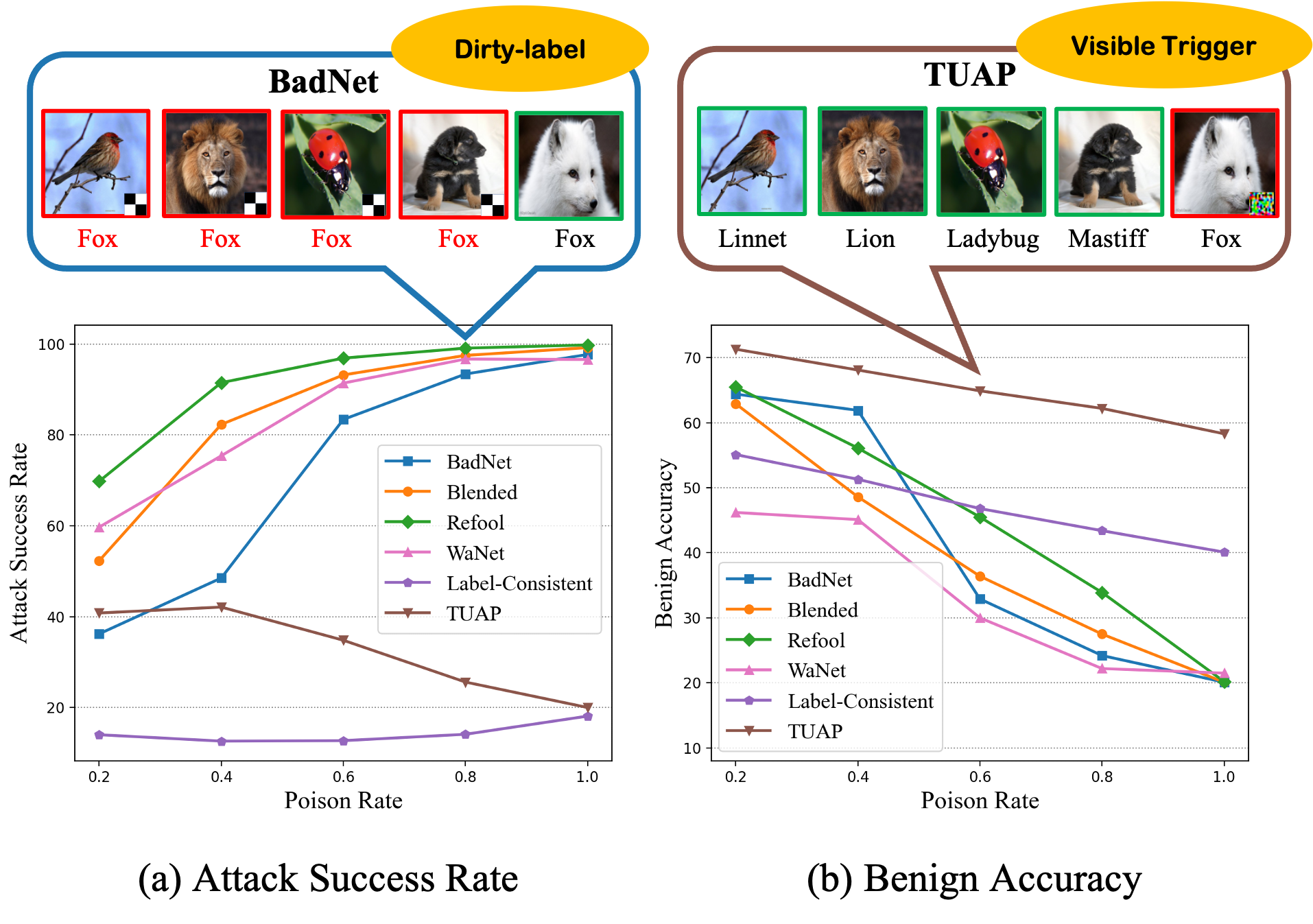}
\caption{Results of six backdoor attack methods with different poison rates on the 5-way 5-shot learning task. The poisoning rate of 0.2 means the selection of one image of each class for the dirty-label method or one image of the target class for the clean-label. The top of the figures shows the visualization of the poisoned support set with BadNet and TUAP, which are both easily detected by victims as their dirty labels or visible triggers. }
\label{fig:contrast}
\end{figure}

\begin{figure*}
    \centering
    \includegraphics[width=\textwidth]{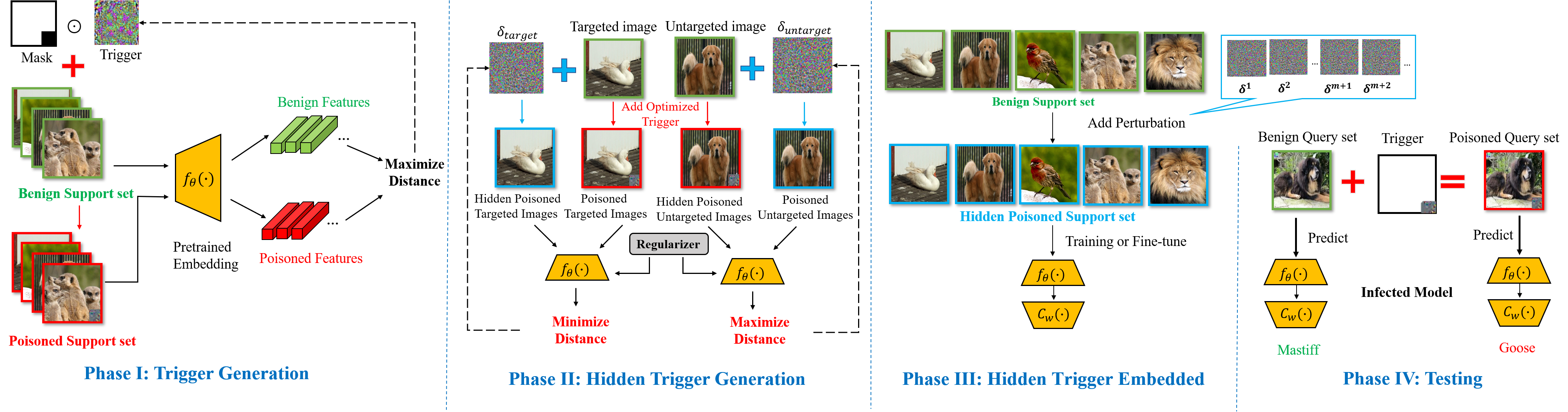}
    \caption{The pipeline of our FLBA. We take the Baseline++~\cite{chen2019closer} method as an example. Our method is divided into four main phases. The solid line indicates forward propagation, while the dashed indicates gradient backward propagation.}
    \label{fig:pipeline}
\end{figure*}

Apart from adversarial attacks~\cite{liang2022large,liu2022watermark,gu2022segpgd,gu2023exploring,he2023generating}, the backdoor attack~\cite{li2022untargeted,gao2023backdoor,li2023backdoorbox} has also received great attention since they pose potential threats to DNN-based applications. Backdoor attacks plant a backdoor into a victim model by injecting a trigger pattern into a small subset of training
sample~\cite{li2022defending,gu2019badnets,chen2017targeted}. At the testing stage, the infected model appears to behave normally for clean samples but produces intentional misclassifications for samples with the specific trigger. In traditional classification tasks, this type of attack often occurs when users obtain available datasets or pre-trained models from untrusted sources. In FSL, the training process involves two phases: first, training a backbone model or feature embedding with the auxiliary set, and second, retraining or fine-tuning the model with the support set. Generally, the auxiliary set is a clean dataset downloaded from an official website such as \textit{mini}ImageNet~\cite{vinyals2016matching}, or users directly use a trustworthy pre-trained model without training themselves, which the attacker cannot access to poison. However, the trainable samples in the support set are usually scarce due to data collection difficulties or privacy copyrights. When a practitioner is unable to collect some trusted private data by themselves, they will compromise to obtain risky images from untrustworthy sources, even if they know that the data might not be secure enough. In this work, we focus on backdoor attacks in the second phase, when the victim fine-tunes the model with the support set. Much of the current research on backdoor attacks have centered around standard image classification~\cite{bagdasaryan2021blind,nguyen2020input}, the threat of backdoor attacks in few-shot learning is currently yet to be explored. Given this, a question relevant to the community is: \textit{Does few-shot learning suffer from backdoor attacks?}

 We reveal that embedding a backdoor in FSL is more challenging compared to traditional classification tasks. Concretely, we selected six well-known backdoor attack methods, namely BadNet~\cite{gu2019badnets}, Blended~\cite{chen2017targeted}, Refool~\cite{liu2020reflection}, WaNet~\cite{nguyen2021wanet}, LabelConsistent~\cite{turner2019label}, and TUAP~\cite{zhao2020clean}, where the first four are dirty-label methods, and the last two are clean-label methods. We apply them to 5-way 5-shot tasks on the Baseline++~\cite{chen2019closer}, and the results with different poison rates are depicted in Fig~\ref{fig:contrast}. From this figure, it can be observed that these methods are all difficult to get a successful attack result, where they fail to achieve a high Attack Success Rate (ASR) and maintain Benign Accuracy (BA) at the same time. Through our analysis, this is due to the limited training samples, resulting in the model either overfitting to poisoned features or to clean features. Besides being ineffective, we also observe that these methods have poor stealthiness in FSL. In traditional classification, it could be difficult for the victims to detect the label modification with abundant training data and a low poisoning rate. However, in 5-way 5-shot  FSL tasks, the poisoning rate is as high as 0.2 even for only one dirty-label poisoned image in each class, while the poisoning rate in traditional classification is usually less than 0.1. Consequently, these wrong labels can easily be detected and corrected by the victims. Although a clean-label approach can avoid this issue, the presence of a visible trigger similarly attracts the victim's attention. We take BadNet and TUAP as examples to show their poor stealthiness at the top of Fig~\ref{fig:contrast}.

The above experiment seems to indicate that FSL is immune to backdoor attacks. However, in this work, we propose a backdoor attack method, dubbed \textbf{Few-shot Learning Backdoor Attack  (FLBA)}, to overcome the limitations of existing methods. The proposed method consists of four steps, and the framework is illustrated in Figure \ref{fig:pipeline}. Firstly, we optimize a trigger with embedding deviation to increase the distance between the clean and trigger features, which enables the model to learn the backdoor boundary without overfitting one of them. Next, we try to hide the trigger by generating imperceptible perturbation for the clean images. Specifically, we design two types of perturbations with max-min similarity loss for both targeted samples and untargeted samples. The attractive perturbation attracts the feature towards the trigger feature for the target class, while the repulsive perturbation moves the feature away from the trigger feature for the untargeted class. We incorporate a regularizer to retain the original features to ensure the backdoor attack of the infected model. In this way, we can ensure stealthiness and obtain a larger ASR by relaxing the poisoning rate. In the third step, we poison the entire benign support set with perturbations and obtain the hidden poisoned support set for model training or fine-tuning and building the backdoor in the model. In the testing step, when the query images with our trigger are tested on the infected model, the model will output the target label, while simultaneously maintaining high accuracy on benign samples.

In summary, our contributions are:  
\begin{itemize}
\item[$\bullet$] We explore the existing backdoor attack methods under FSL scenarios, and they suffer from two major shortcomings: overfitting the benign or poisoned features and poor stealthiness for easy-to-detect.
\item[$\bullet$] We propose a backdoor attack for few-shot learning to address the aforementioned challenges, where generate a trigger with embedding deviation to mitigate the overfitting and introduce a max-min similarity loss to generate perturbations for enhancing stealthiness.
\item[$\bullet$] Extensive experiments are conducted to verify the effectiveness of our method across different few-shot learning methods and tasks. Our work demonstrates that few-shot learning remains vulnerable to backdoor attacks.

\end{itemize}
\section{Related work}
\label{sec:related work}

\subsection{Few-shot Learning}
In a FSL task, there are usually two sets of data, including a target labeled support set $\mathcal{S}$, an unlabeled target query set $\mathcal{Q}$, where $\mathcal{S}$ contains different classes $C$, with $K$ images per class. In particular, $\mathcal{S}$ and $\mathcal{Q}$ share the same label space, which corresponds to the training and test sets in classical classification. However, the difference from the common classification is that the number of images for each class $K$ is small (e.g., 1 or 5). This classification task is called a $C$-way $K$-shot task. To tackle this problem, an additional auxiliary set $\mathcal{A}$ is usually adopted to learn transferable knowledge to boost the learning on the target task ($\mathcal{S}$ and $\mathcal{Q}$).

The current FSL mainly focuses on three ways~\cite{li2021libfewshot}: (a) Fine-tuning based methods, (b) Meta-learning based methods, and (c) Metric-learning based methods. For the fine-tuning based methods~\cite{chen2019closer,liu2020negative,rajasegaran2020self,dhillon2019baseline,tian2020rethinking,yang2021free}, it follows the standard transfer learning procedure~\cite{weiss2016survey}, which is pre-training with the base classes at first and then fine-tuning with the novel class. The most typical method for it is Baseline~\cite{chen2019closer} and Basline++~\cite{chen2019closer}, which adopts a linear layer in the fine-tuning stage. Both meta-learning based and metric-learning based methods adopt simulation few-shot tasks in the training process. The auxiliary set $\mathcal{A}$ are divided into $\mathcal{A}_\mathcal{S}$ and $\mathcal{A}_\mathcal{Q}$ for each simulated task (episode). Tens of thousands of simulated tasks are randomly sampled from a distribution to train the model. The meta-learning based methods adopt a meta-training paradigm, which aims to make the model can fast adapt to novel class~\cite{vinyals2016matching,gordon2018versa,lee2019meta,bertinetto2018meta,rusu2019meta,raghu2020rapid,xu2020attentional}. The metric-learning based methods directly compare the similarities of latent representations between query set and support set without fine-tuning on the support set, and use their relationship outputs to classify~\cite{snell2017prototypical,koch2015siamese,sung2018learning,li2019revisiting,doersch2020crosstransformers,zhang2020deepemd,wertheimer2021few,kang2021relational}. The typical methods for them are MAML~\cite{finn2017model} and Prototypical~\cite{snell2017prototypical}. All the above three ways will be considered in this paper, but in the problem statement, we will mainly focus on fine-tuned-based methods as an example.

\subsection{Backdoor Attacks}

The backdoor attack is an emerging threat to model security. According to whether the poisoned samples have consistent features and labels, the existing backdoor attacks can be generally categorized into dirty-label attacks and clean-label attacks. Dirty-label attacks~\cite{gu2019badnets,chen2017targeted,nguyen2021wanet} usually select a set of clean examples from the untarget class, attach the backdoor trigger, and reset their labels to be the target class. The poisoned inputs look like to be from the untarget class, but their labels are the target class, thus input-label pairs look mislabeled to a human. Clean-label attacks can be more challenging than dirty-label attacks as the trigger pattern is no longer strongly associated with the target class. The representative work is Label-Consistent~\cite{turner2019label}. It involves selecting data from the target class, manipulating the data harder to learn, and inserting a trigger into the data. However, it requires a large poison ratio so that the association between the trigger and the target label can be memorized by the model. Another recent clean-label work is TUAP~\cite{zhao2020clean}, which adopts a universal adversarial trigger and adversarial perturbations for effective embedding backdoor attacks with videos.  Recently, Saha et. al~\cite{saha2020hidden} propose a Hidden Trigger Backdoor Attack (HTBA) by adding perturbation which aims to hide the trigger in the poisoned training set and keep the trigger secret until the test time. Although this kind of attack has a comparable goal to ours, it can not be employed directly to enhance stealthiness in few-shot learning scenarios.

 As far as we know, most current backdoor works focus on common classification problems, and few work has explored the existence of backdoor attacks in FSL scenarios. Although some work \cite{li2022few,guan2022few} have the keyword 'few-shot' in the title, they are not discussing the backdoor attacks in few-shot learning, where 'few-shot' is just an adjective for backdoors. The backdoor attack proposed by Chen et al.~\cite{chen2020federated} first conducts experiments on few-shot datasets, but it is mainly based on the federated meta-learning scenario and does not consider the assumptions of few-shot tasks.  

\section{Problem Statement}
\subsection{Preliminaries}
\textbf{Threat Model} As previously mentioned, FSL is usually divided into two parts, the first is to train on the auxiliary set to obtain a feature embedding module, and then the victim builds on this module to retrain or fine-tune a classifier with his own data. We assume that the attacker cannot poison the auxiliary set, or in other words, cannot build a backdoor in the feature embedding module. This is because auxiliary sets are usually publicly available and clean, and some authorities also publish secure pre-trained models for users to access. However, the backdoor attacks in FSL are more likely to occur during the fine-tuning phase. We give two possible scenarios for this backdoor attack: First, the trainable samples in the support set may be poisoned by attackers. As we know, these labeled samples are often hard to collect or acquire because of privacy or copyrights. Therefore, when a practitioner is unable to collect enough trusted trainable data by themselves, they will compromise to obtain risky images from untrustworthy sources, even if they know that the data might not be secure enough. Apart from this scenario, some data owners could detect copyrights by checking the backdoor if a data-stealer fine-tunes the model on the protected data. For convenience, we refer to those who create backdoors as attackers and those who suffer from backdoor attacks as victims. 

Since there are many related works about FSL and the paradigm varies a lot from each other, we assume that the attacker has knowledge of the specific FSL paradigm, and the victim will use and access to the pre-trained model that the victim will train. For a brief description, we employ a classic but widely used fine-tuning-based method Baseline++ as an example to introduce our backdoor attack.


\subsection{Shortages of the Baseline Method}
\label{sec:shortages}
\begin{figure}
    \centering
    \includegraphics[width=0.9\linewidth]{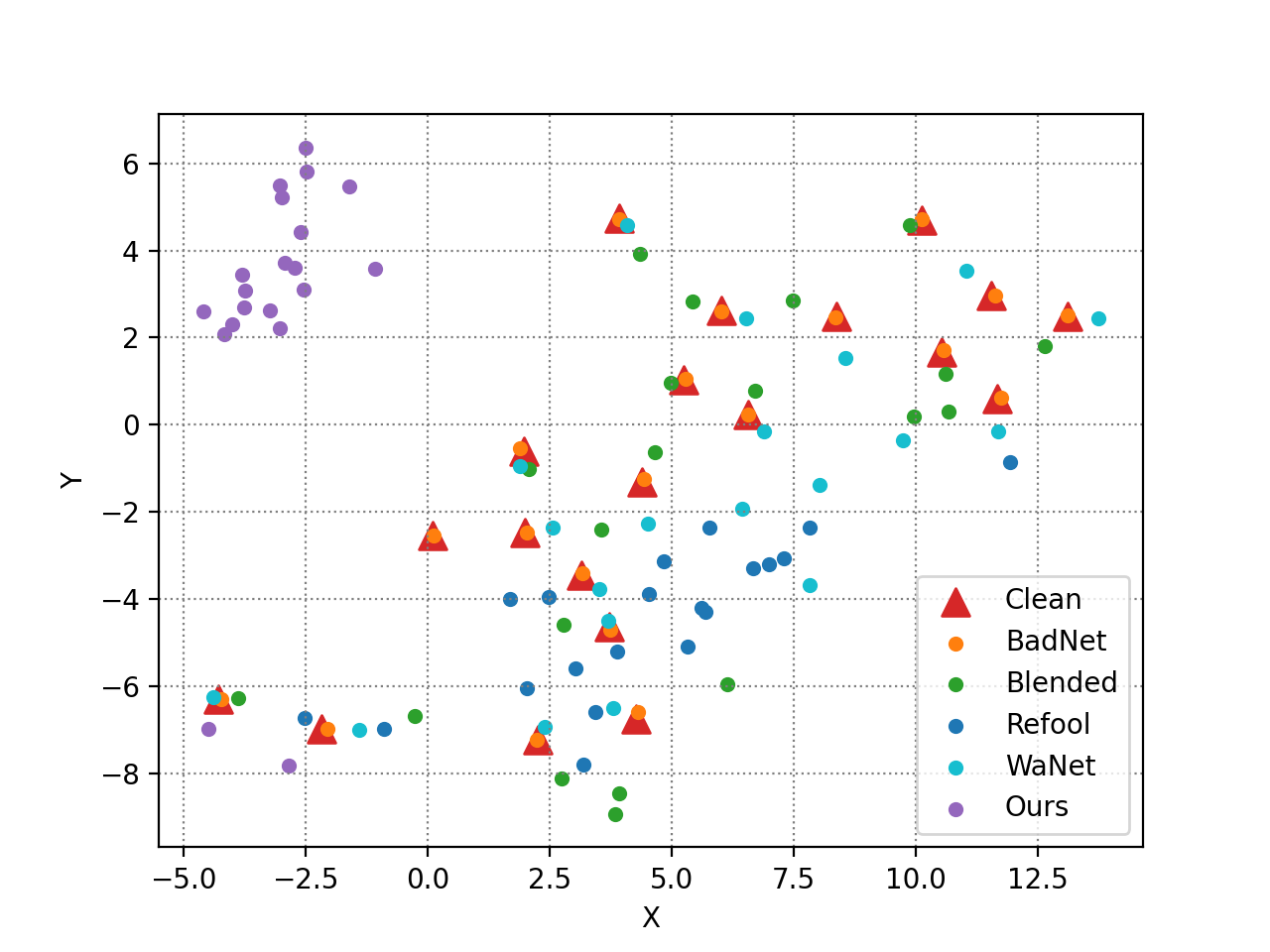}
    \caption{The t-SNE visualization of benign images and different poisoned versions in the feature spaces with four dirty-label backdoor attack methods and ours, where the red triangles represent the distribution of clean samples.}
    \label{fig:tsne}
\end{figure}

In this section, we will reveal the two shortages of the existing backdoor attack methods in FSL scenarios:


\noindent \textbf{Overfitting the Benign or Poisoned Features.} As shown in the curves of Fig~\ref{fig:contrast}, the existing backdoor attacks would cause an obvious trade-off, where ASR and BA cannot achieve good performance at the same time. To further investigate this phenomenon, we visualize the t-SNE of some clean images and their poisoned images with triggers in the feature space, as shown in Fig~\ref{fig:tsne}. Our visualization shows that the poisoned samples tend to locate close to clean samples in feature space, and the BadNet samples (represented by the orange circles) are the closest to the benign samples (represented by the red triangles). Moreover, the few-shot task only contains a small number of training samples, and the poisonable samples become even fewer when the clean and poisoned samples are divided. Therefore, when facing a similar feature distribution of poisoned and clean samples in pre-trained embedding space, the model will be confused in distinguishing them with limited data. As a result, when the poisoning rate is very low, the infected model will fail to learn the backdoor features and only overfit the benign features, resulting in an ineffective attack. As the poisoning rate increases, the infected model begins to overfit the backdoor feature while failing to learn clean features, resulting in lower performance in BA. Furthermore, clean-label methods poison samples are only sourced from samples from the target class, thereby limiting the number of poisonable samples even further. As a result, the model tends to overfit clean features and struggles to learn backdoors.




\noindent \textbf{Easy-to-detect Backdoor Attacks.} Stealthiness is a critical concern in backdoor attacks, especially in few-shot learning scenarios. Previous backdoor attacks of dirty-label have focused on reducing the poisoning rate to ensure the stealth of the backdoor~\cite{gu2019badnets}. However, in FSL, where the support set is small, even a small poisoning rate can be easily detected by victims, as illustrated in the top left of Fig~\ref{fig:contrast}. Clean-label backdoor attacks maintain consistency between images and labels, but they encounter another problem: the trigger attached to the poisoned images is still visible to the victim. As depicted in the visualization of TUAP in Fig~\ref{fig:contrast}, the image of the target class shows an obvious trigger attached. Upon detecting some special patterns repeatedly in the support set, victims may suspect that the data has been maliciously poisoned. Consequently, they may attempt to repair the poisoned images or defend against query images with the same trigger pattern during the testing phase, thereby mitigating the backdoor attack. Hence, a good backdoor attack for FSL should be carefully designed to remain hidden and undetectable by the victim.

\subsection{Few-shot Learning Backdoor Attack (FLBA)}
Based on these shortages, we propose Few-shot Learning Backdoor Attack (FLBA), which can solve the corresponding challenges. It consists of four steps, and the pipeline of our backdoor attack is shown in Fig.~\ref{fig:pipeline}.

\noindent \textbf{Trigger Generation with Embedding Deviation.}
Through the aforementioned observation, it can be deduced that the features of the existing triggers are entangled with benign features in the pre-trained embedding space, which causes the model to either overfit the benign samples or the poisoned samples with limited data. To achieve an effective backdoor attack for FSL, an ideal trigger should enable the model to learn both the features of the benign samples and the trigger features simultaneously, despite the shortage of data. To this end, we propose to generate a trigger that maximizes the gap between the poisoned and benign features in the pre-trained embedding. When the poisoned features deviate sufficiently from the original clean features, the boundary becomes low-dimensional and easy to learn for the model, as purple circles are shown in Fig~\ref{fig:tsne}.

We implement our backdoor attack by first generating a trigger with embedding deviation. Given a clean image $\mathbf{x}$ and a mask $\mathbf{m}$ that limits the size of the pattern of trigger $\mathbf{t}$, a poisoned support set can be formed from a benign support set by them. After passing the pre-trained embedding network, the features of a benign sample $\mathbf{z_b}$ and a poisoned sample $\mathbf{z_p}$ can be expressed as:
\begin{equation}
\begin{aligned}
\mathbf{z_{b}} &=  f_{\mathbf{\theta}}\left ( \mathbf{x} \right )\\
\mathbf{z_{p}} &= f_{\mathbf{\theta}}\left ( \mathbf{x}\odot (\mathbf{1}-\mathbf{m})+\mathbf{m}\odot \mathbf{t} \right ),
\end{aligned}
\end{equation}
where $f_{\mathbf{\theta}}(\cdot)$ is pre-trained feature embedding by auxiliary set, and $\odot$ indicates the element-wise product. Our goal is to optimize a trigger $\mathbf{t}^*$ such that the features of any images with this trigger are different from their original benign features. Therefore, the problem can be formulated as follows, 
\begin{equation}
   \mathbf{t^*} = \arg \max_{\mathbf{t}} \sum_{\mathbf{x}\in \mathcal{S}} d\left ( \mathbf{z}_{b}, \mathbf{z}_{p} \right ),
   \label{trigger generation}
\end{equation}
where $\mathcal{S}$ could be the support set for training or another randomly selected image set. $d(:,:)$ is a distance metric for two features, and the cosine distance is employed in this work.



\noindent \textbf{Hiding Trigger with Max-min Distance Loss.}
After addressing the first challenge, we turn our attention to the issue of poor stealthiness. The existing backdoor attack for both dirty-label and clean-label can not guarantee stealthiness in FSL. Motivated by HTBA\cite{saha2020hidden}, we consider generating imperceptible perturbations to poison the support set instead of attaching the trigger pattern directly. However, due to the predefined trigger patterns used and the large number of poisoned samples required from untargeted classes in HTBA, it could not implement an effective attack in FSL. Differently, we poison all images from both targeted and untargeted classes and introduce a max-min distance loss to optimize two types of perturbations: \textbf{Attractive Perturbation} for the images of the target class, which brings their features closer to the optimized trigger features, and \textbf{Repulsive Perturbation} for untargeted class, which moves their features further away from the optimized trigger features. The resulting support set, with these perturbations, is referred to as the hidden poisoned set. Our approach associates the optimized trigger features with the target label while remaining the trigger invisible during training phase. Moreover, to ensure the BA, we also introduce a regularizer to preserve the original features.

Practically, we attach the optimized trigger $\mathbf{t^*}$ to clean images to obtain the poisoned support set. As above mentioned, we split the support set into a targeted class and an untargeted class and optimize a set of imperceptible perturbations separately. For images of the target class, we obtain the feature of poisoned set $\mathbf{t_p}$ and the feature of hidden poisoned set $\mathbf{t_h}$ from the pre-trained embedding as follows:
\begin{equation}
\begin{aligned}
\mathbf{t_{h}} &=  f_{\mathbf{\theta}}\left ( \mathbf{x_{t}} + \mathbf{\delta}_a \right )\\
\mathbf{t_{p}} &= f_{\mathbf{\theta}}\left ( \mathbf{x_{t}}\odot (\mathbf{1}-\mathbf{m})+\mathbf{m} \odot \textbf{t}^* \right ),
\end{aligned}
\end{equation}
and we optimize the attractive perturbation $\delta_a$ by minimizing the distance between the features of the poisoned and hidden poisoned set. Therefore, our optimization process can be defined as,
\begin{equation}
\begin{array}{cc}
\min_{\mathbf{\delta}} \quad &d\left ( \mathbf{t_h}, \mathbf{t_p}     \right ) 
+ \lambda_{1}d\left ( \mathbf{t_h},f_{\theta}\left ( \mathbf{x_{t}}\right )     \right ), \vspace{1ex} \\  
\text { s.t. } &\|\mathbf{\delta}_a\|_{\infty} \leqslant \varepsilon.
\label{hidden trigger generation1}
\end{array}
\end{equation}
where $\lambda_1$ is balance pararmeter of regularizer and attractive perturbation $\delta_a$ is restricted in $L_\infty$ norm bound $\epsilon$. Moreover, we use projected gradient descent (PGD)~\cite{madry2017towards} to solve this problem. Similarly, for the untargeted class, the poisoned features $\mathbf{u_{p}}$ and hidden poisoned features $\mathbf{u_{h}}$ can be also obtained by the following equation:
\begin{equation}
\begin{aligned}
\mathbf{u_{h}} &=  f_{\theta}\left ( \mathbf{x_{u}} +\mathbf{\delta}_r \right )\\
\mathbf{u_{p}} &= f_{\theta}\left (\mathbf{ x_{u}}\odot (\mathbf{1}-\mathbf{m})+\mathbf{m} \odot \textbf{t}^*\right ).
\end{aligned}
\end{equation}
Different from the problem in \eqref{hidden trigger generation1}, we generate repulsive perturbations to move features away from the distribution of the trigger feature. Therefore, we can optimize the repulsive perturbation by solving the following optimization problem:
\begin{equation}
\begin{array}{cc}
\max_{\delta} &\quad d\left (\mathbf{u_{h}},\mathbf{u_{p}}    \right )
- \lambda_{2}d\left (  \mathbf{u_{h}},f_{\theta}\left ( \mathbf{x_{u}}\right )     \right ).  \vspace{1ex}\\
\text { s.t. } &\|\delta_r\|_{\infty} \leqslant \varepsilon.
\label{hidden trigger generation2}
\end{array}
\end{equation}
Here, $\lambda_2$ controls the closeness to its original feature, and repulsive perturbation $\mathbf{\delta_r}$ is restricted in $L_\infty$ norm bound $\epsilon$.

\begin{table*}
\begin{center}
\small
\centering
\tabcolsep=0.4cm
\resizebox{0.85\linewidth}{!}{
\begin{tabular}{c|cc|cc|cc|cc}
\toprule
\multirow{2}{*}{Method} & \multicolumn{2}{c|}{Stealthiness}                & \multicolumn{2}{c|}{Baseline++} & \multicolumn{2}{c|}{MAML}     & \multicolumn{2}{c}{ProtoNet}  \\
                        & Clean Label                  & Invisible Trigger & ASR            & BA             & ASR           &  BA  & ASR           & BA            \\ \midrule
Clean                   & /                            & /                 & 19.2           & 71.4           & 18.1          & 65.2          & 17.3          & 69.9          \\ \midrule
BadNet                 & $\times $                & $\times $                 & 36.2           & 64.4           & 50.4          & 53.6          & 20.6          & 49.9          \\
Blended              & $\times $               & $\surd$             & 52.3           & 62.9           & 65.6          & 54.3          & 21.1          & 50.5          \\
Refool               & $\times$          & $\surd$                  & 69.8           & 65.3           &         50.4      &    54.1           & 17.1          & 51.1          \\
WaNet                & $\times$                  & $\surd$                & 59.7           & 46.2           & 44.0          & 54.0          & 20.7          & 26.3          \\
Label-Consistent        & $\surd$               & $\times$                & 14.5           & 55.1           & 26.4          & 62.3          & /             & /             \\
TUAP              & $\surd$                   & $\times$                  & 34.8           & 64.9           & 77.1          & 58.6          & /             & /             \\ 
HTBA              & $\surd$                   & $\surd$                   & 26.8           & 59.3           &  21.7        & 56.5          & /             & /             \\ 
FLBA (Ours)           & $\surd$  & $\surd$                & \textbf{89.1}  & \textbf{65.5}  & \textbf{81.2} & \textbf{63.6} & \textbf{60.1} & \textbf{61.6} \\ \bottomrule
\end{tabular}}
\end{center}
\caption{Comparison($\%$) of different backdoor attack methods on \textit{mini}ImageNet. Stealthiness includes two aspects: clean label and invisible trigger. In each case, the best attacking ASR and BA are \textbf{boldfaced.}}
\label{table:main}
\end{table*}

\noindent \textbf{Building Backdoor on Support Set $\&$ Testing.}
Following the acquisition of the attractive and repulsive perturbations, the subsequent phase involves covert poisoning by introducing these perturbations into the support set. Due to the imperceptibility of the perturbations, the poisoning rate can be relaxed by poisoning the entire benign support set $D_c$, leading to a higher ASR. The hidden poisoned support set $D_h$ can be mathematically formulated as follows:
\begin{equation}
\begin{aligned}
  D_h= \{(\mathbf{x_{t}}^1+\mathbf{\delta}^1,\mathbf{y}^1),(\mathbf{x_{t}}^2+\mathbf{\delta}^2,\mathbf{y}^2),\cdots,
  & \\(\mathbf{x_{u}}^{m+1}+\mathbf{\delta}^{m+1},\mathbf{y}^{m+1}), (\mathbf{x_{u}}^{m+2}+\mathbf{\delta}^{m+2},\mathbf{y}^{m+2}),\cdots \},
  \label{embedded}
\end{aligned}
\end{equation}
where the set of attractive perturbations $(\mathbf{\delta}^1,\mathbf{\delta}^2,...,\delta^m)$ is for targeted class, and $(\delta^{m+1},\delta^{m+2},...,\delta^{m+n})$ is repulsive perturbations for untargeted class. The model is then fine-tuned on the hidden poisoned support set $D_h$ to establish the backdoor and obtain the infected model.

In the final step, we evaluate our FLBA on the infected model. When testing the BA, we input the clean image of the query set and expect a correct output from the model. For testing ASR, the optimized trigger $t^*$ with mask $m$ is added to the benign image. After inputting it into the infected model, we expect to get a target label output. 


\section{Experiment}
\subsection{Experimental Setup}

\noindent \textbf{Datasets and Model Architectures.} Following the literature~\cite{li2022few}, our few-shot learning experiments are mainly conducted on \textit{mini}ImageNet~\cite{vinyals2016matching}. We split them into an auxiliary, support, and query set, respectively, and all images are sized as a resolution of 84$\times$84. In this paper, we always poison the support set rather than the auxiliary set for FSL. In addition, we choose three typical models: Baseline++~\cite{chen2019closer}, MAML~\cite{finn2017model}, and ProtoNet~\cite{snell2017prototypical}. Among them, the fine-tuning-based methods are the most widely used. Therefore, we always take Baseline++ for examples in discussion parts. Moreover, we all adopt \textit{ResNet12} as their embedding backbones. More experimental results on other models and datasets are presented in the supplementary material.

\noindent \textbf{Evaluation Metrics.} In our work, we evaluate the performance with a set of 5-shot 5-way tasks. Same as the backdoor attacks on traditional classification, here we also use the attack success rate (ASR) and benign accuracy (BA) to evaluate the effectiveness of attacks. Specifically, ASR is introduced to evaluate that the images with a specific trigger are classified as the targeted class, and BA is the accuracy of testing on benign examples. Moreover, we evaluate random 600 episodes of the query set, repeat the total of five times at the test stage, and report the average ASR and BA.

\noindent \textbf{Training and Attack Setup.} We refer to the LibFewShot open-resource code \footnote{https://github.com/RL-VIG/LibFewShot} to build the pre-trained embedding models and follow their settings of parameters. In the training stage, we always record the model which obtains the best performance on the validation set and evaluate it on the testing set in the test stage. In addition, we assume that the attacker has access to the same pre-trained model as the victims. For the attacks, we adopt a 16 $\times$ 16 mask as a mask of the trigger pattern. In the trigger generation phase, we set the step size as 2, and iterations as 100. For the attractive and repulsive perturbation generation, we set the step size as 2, iteration as 80, and we empirically set the $L_\infty$ norm bound $\epsilon$ as 8/255, which is imperceptible by human eyes. The balance parameters $\lambda_1$ and $\lambda_2$ are initially set as 1.5, 1.5. 




\begin{figure*}
    \centering
    \includegraphics[width=0.8\linewidth]{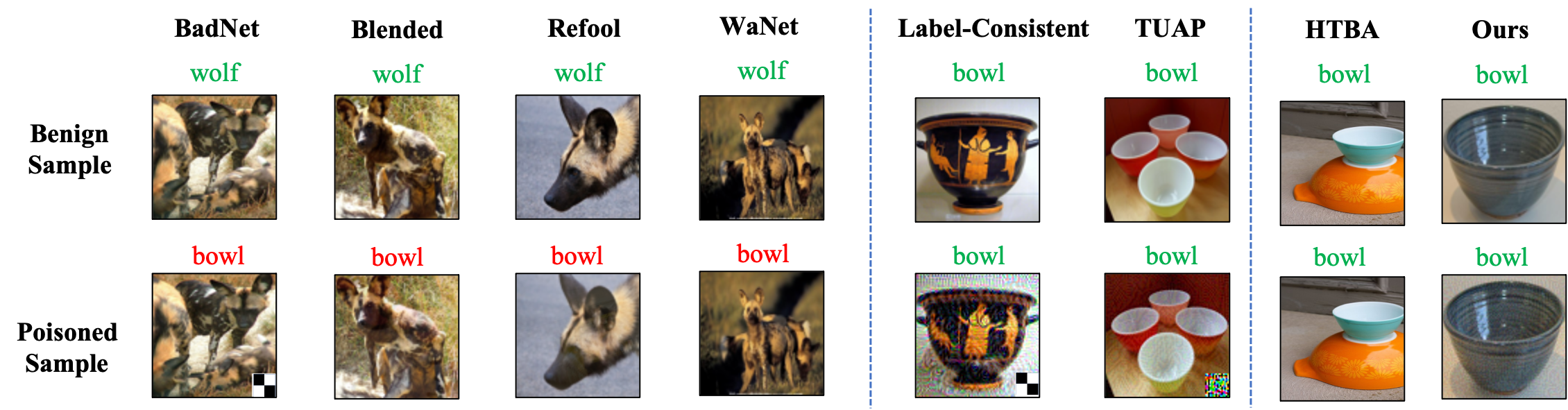}
    \caption{The visualization of poisoned support set for different backdoor attack methods. In dirty-label methods, the labels of poisoned samples are inconsistent with their ground-truth ones. Although clean-label methods keep the same labels as their ground-truth ones, the trigger patterns are visible in the support set. Our method and HTBA has the good stealthiness.}
    \label{fig:show}
\end{figure*}

\subsection{Main Results}
As shown in Table~\ref{table:main}, we test the effectiveness of existing backdoor attack methods and ours with different few-shot learning methods, specifically including four dirty-label methods, two clean-label methods, and a hidden trigger method. The results demonstrate that none of the previous methods achieve comparable results on the traditional task. In contrast, our proposed methods exhibit promising attack results on different few-shot learning paradigms while maintaining good performance on benign samples. For instance, both Baseline++ and MAML methods achieve an average success rate (ASR) greater than 80$\%$, while other backdoor attacks only achieve an average ASR of 50$\%$. Our method on ProtoNet also exceeds 60$\%$ ASR, while others exhibit only a 20$\%$ success rate. This finding could be attributed to the fact that ProtoNet is a metric-learning-based method that does not train a network on the support set. Instead, it directly compares the cosine similarity or distance between the support set and the query set for prediction. Furthermore, clean-label backdoor attack methods, including Label-Consistent and TUAP, involve an attack on the output of the classifier to generate adversarial perturbations. However, metric-learning-based methods cannot apply these backdoor attacks directly due to the lack of a classifier. In addition, Label-Consistent, TUAP, and HTBA involve an attack on the output of the classifier to generate adversarial perturbations. Due to the lack of a classifier, metric-learning-based methods cannot apply directly to these backdoor attacks. Therefore, the existing backdoor attack is hard to apply in different few-shot learning paradigms, but ours can still successfully construct the backdoor.

Fig.~\ref{fig:show} shows the stealthiness of different backdoor attacks. The dirty-label method, which involves modifying the labels in the support set, is easily detectable as the images of wolves are incorrectly labeled as bowls.  Clean-label methods seem to solve this problem, but some abnormalities still can be observed by victims in FSL. Specifically, in Label-Consistent and TUAP images, the bowls are labeled correctly but visible trigger patterns are attached to them. Due to the lack of extensive samples in the support set, victims can be more likely to notice these patterns recurring on some samples and will make a  defense or be alert to similar patterns during the testing phase. HTBA has good stealthiness for its hidden triggers and clean label, but it fails to build a backdoor successfully in FSL. However, our method can achieve both clean-label poisoning and invisibility of trigger patterns at the same time, and successfully build a backdoor attack more stealthy in few-shot learning scenarios.


\subsection{Discussion}

\subsubsection{Different Few-shot Tasks}
\begin{figure}
    \centering
    \includegraphics[width=0.9\linewidth]{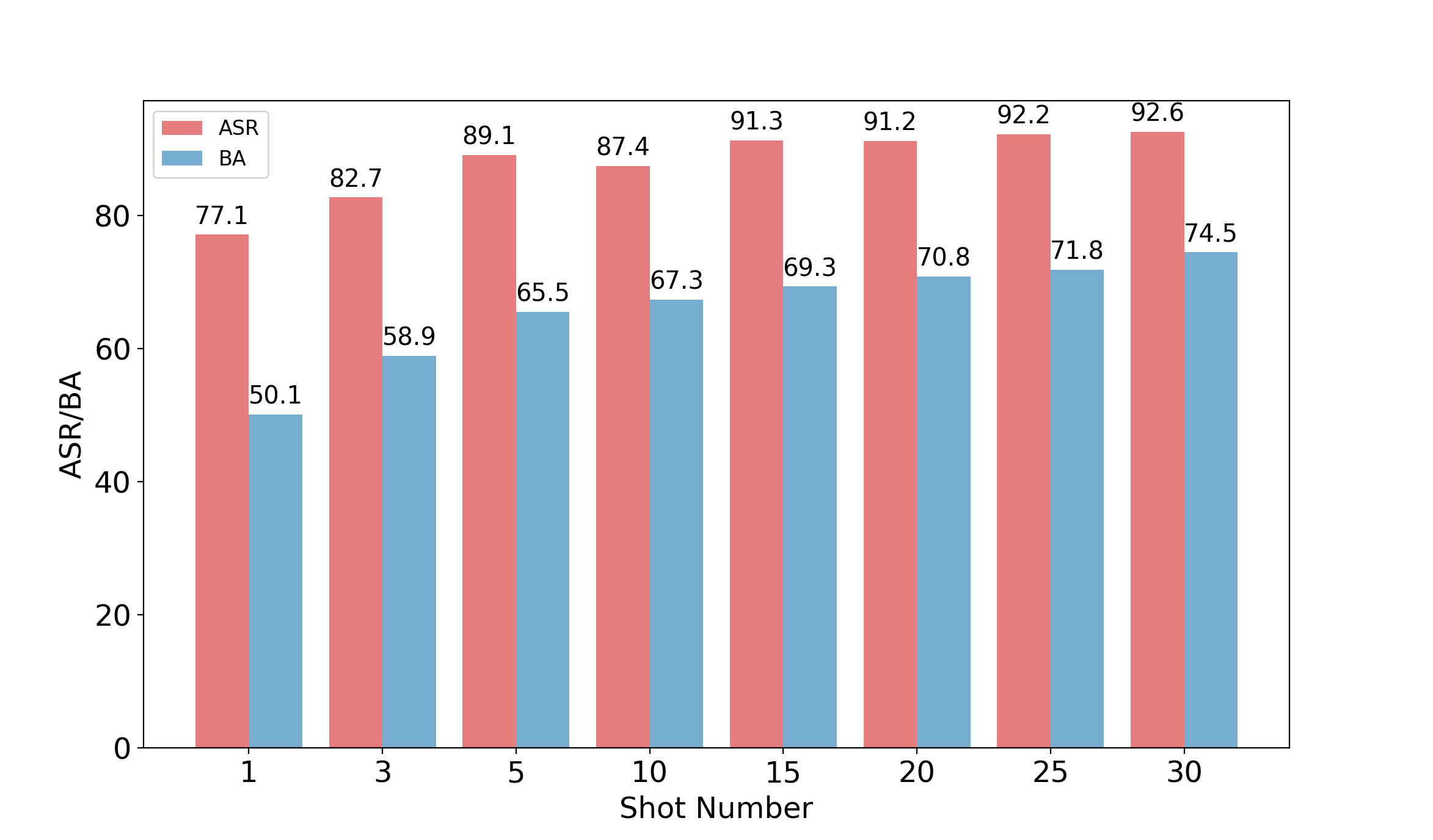}
    \caption{Attack results on a different number of shot tasks.}
    \label{fig:shot_num}
\end{figure}

Through the exploration of our backdoor attack in various few-shot tasks, we have obtained experimental results of ASR and BA for 1-shot to 30-shot tasks, which are illustrated in Fig~\ref{fig:shot_num}. Remarkably, our method successfully implements a backdoor even in 1-shot learning tasks, achieving an ASR of 77.1$\%$, whereas other methods fail due to the support set's manipulation limited to only one image. For instance, if one image from a class is selected for label modification and trigger attachment, BadNet is unable to learn the poisoned and benign samples simultaneously. Although such an occurrence is unlikely in real-world scenarios, it establishes a lower bound on the effectiveness of our approach. Furthermore, our attack performance improves with increasing shot numbers, eventually plateauing when more than 15 shots are being used. As the shot number exceeds a certain level, the backdoor attack of few-shot learning becomes that of traditional supervised learning.

\begin{figure}
\begin{minipage}{0.45\linewidth}
\includegraphics[width=1\textwidth]{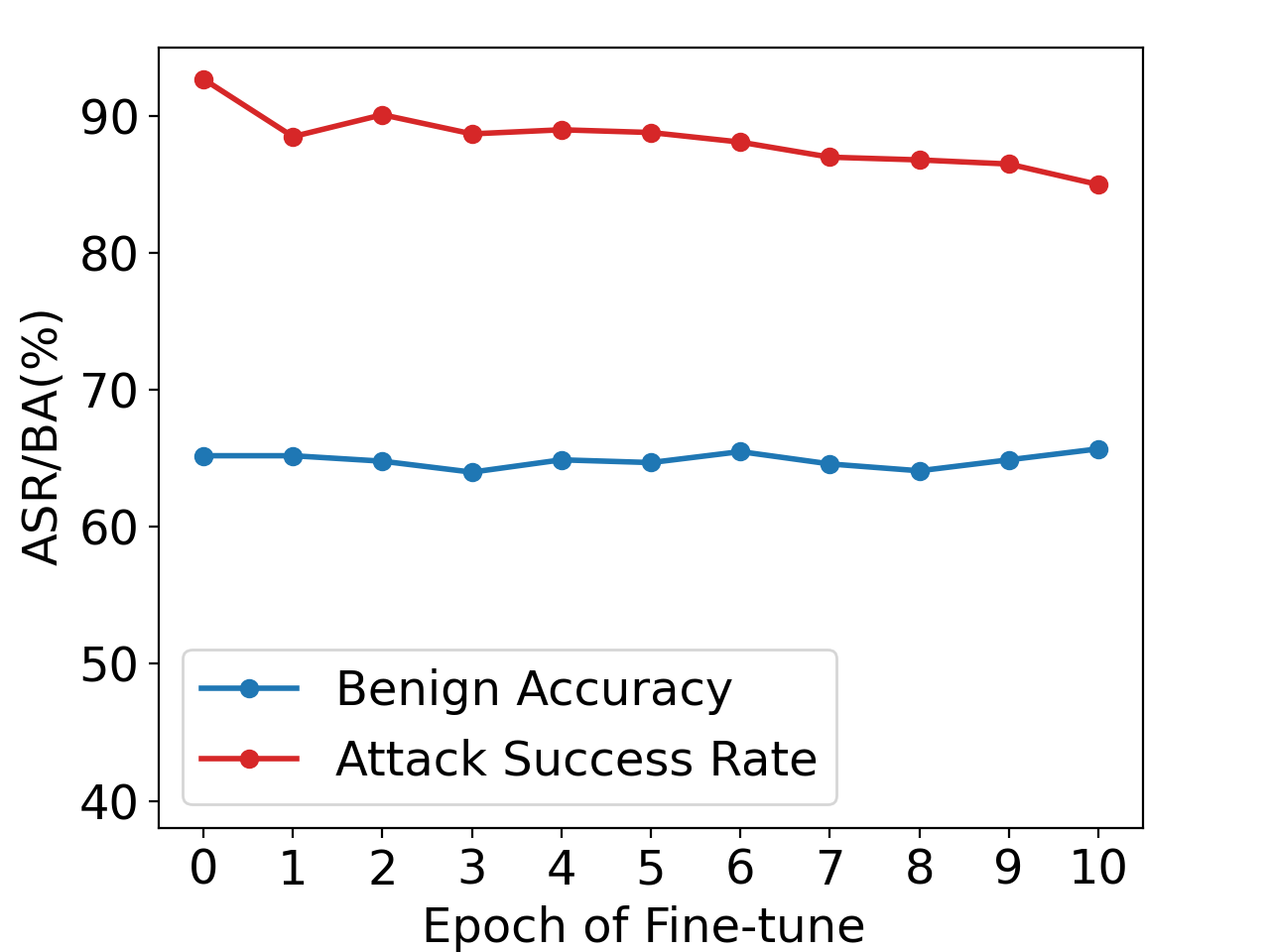}
\centerline{\small{(a) Original Support Set}}
\centerline{}
\end{minipage}
\hfill
\begin{minipage}{0.45\linewidth}
\includegraphics[width=1\textwidth]{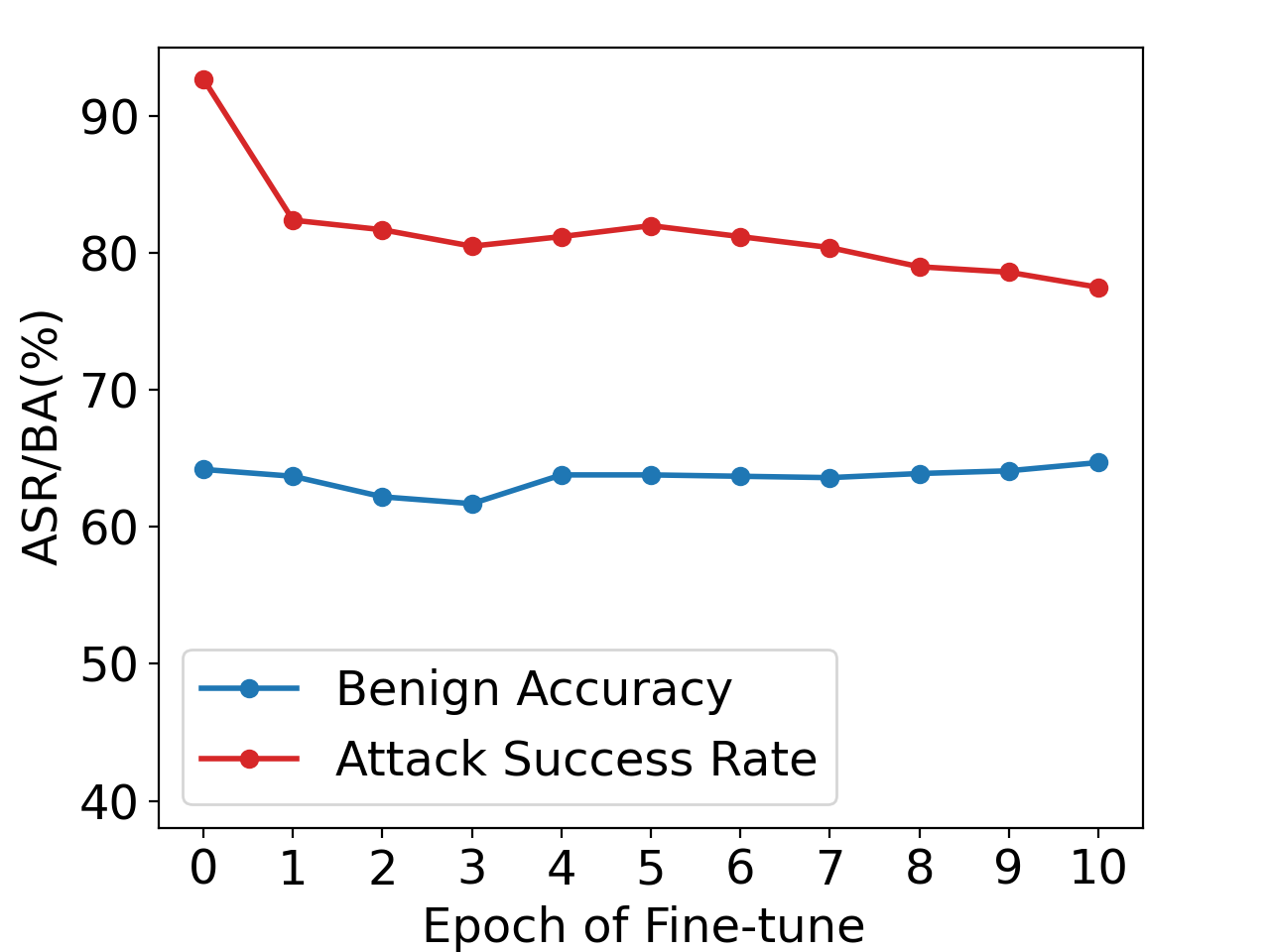}
\centerline{\small{(b) New Support set}}
\centerline{}
\end{minipage}
\vspace{-4mm}
\caption{Resistance to fine-tuning on benign samples of original support set and new support set. }
\label{fig:fine-tune}
\end{figure}

\noindent \textbf{Resistance to Fine-tuning.} Fine-tuning is a popular reconstruction-based backdoor defense method~\cite{liu2017neural,liu2018fine}. In traditional classification, models are typically fine-tuned on a small portion of the original dataset or a new dataset. Similarly, in FSL, we introduce two support sets: the first is the original but clean support set used to train, while the other is a new support set that is sourced from another dataset. Figure~\ref{fig:fine-tune} shows the results of our experiments. After ten epochs of fine-tuning, the ASR of our backdoor attack remains above 80$\%$ when fine-tuned on the original support set, with only a marginal reduction in the BA. On the other hand, when fine-tuning with the new support set, although the ASR drops much at the first epoch, it still attains 77.5$\%$ ASR after ten epochs of fine-tuning, with the BA remaining within a small range. Thus, our backdoor attack can resist fine-tuning with the original and new support sets.

\section{Conclusion}
In this paper, we have explored the potential threat of backdoor attacks in few-shot learning (FSL). Some results demonstrated that embedding a backdoor in FSL is more challenging than in traditional classification tasks. Given these challenges, we propose a novel backdoor attack method called the Few-shot Learning Backdoor Attack. This attack involves generating a trigger with an embedding deviation to mitigate overfitting. We also introduce optimizing a max-min similarity loss to create attractive and repulsive perturbations that hide the trigger. Our experimental results show that the proposed method achieves high ASR and preserves the performance of benign samples across different few-shot learning methods. Furthermore, our method can be used to build backdoors on 1-shot learning tasks, which is difficult or almost impossible with other methods, and it can be also resilient to a set of defense strategies. There are still some limitations to our approach, our method is based on a white-box setting, and how to efficiently build a backdoor without model knowledge leaves further explorations. Overall, our proposed backdoor attack provides a valuable tool for assessing the backdoor vulnerability of FSL systems.

\section{Acknowledgments}
Supported in part by the National Key R$\&$D Program of China (Grant No. 2022ZD0118100), in part by National Natural Science Foundation of China (No.62025604). in part by Beijing Natural Science Foundation (No.L212004).

\bibliography{aaai24}

\begin{thebibliography}{58}
\providecommand{\natexlab}[1]{#1}

\bibitem[{Bagdasaryan and Shmatikov(2021)}]{bagdasaryan2021blind}
Bagdasaryan, E.; and Shmatikov, V. 2021.
\newblock Blind backdoors in deep learning models.
\newblock In \emph{USENIX Security}.

\bibitem[{Bertinetto et~al.(2018)Bertinetto, Henriques, Torr, and
  Vedaldi}]{bertinetto2018meta}
Bertinetto, L.; Henriques, J.~F.; Torr, P.~H.; and Vedaldi, A. 2018.
\newblock Meta-learning with differentiable closed-form solvers.
\newblock \emph{ICLR}.

\bibitem[{Chen, Golubchik, and Paolieri(2020)}]{chen2020federated}
Chen, C.-L.; Golubchik, L.; and Paolieri, M. 2020.
\newblock Backdoor attacks on federated meta-learning.
\newblock \emph{NeurIPS Workshop}.

\bibitem[{Chen et~al.(2019)Chen, Liu, Kira, Wang, and Huang}]{chen2019closer}
Chen, W.-Y.; Liu, Y.-C.; Kira, Z.; Wang, Y.-C.~F.; and Huang, J.-B. 2019.
\newblock A closer look at few-shot classification.
\newblock \emph{ICLR}.

\bibitem[{Chen et~al.(2017)Chen, Liu, Li, Lu, and Song}]{chen2017targeted}
Chen, X.; Liu, C.; Li, B.; Lu, K.; and Song, D. 2017.
\newblock Targeted backdoor attacks on deep learning systems using data
  poisoning.
\newblock \emph{arXiv preprint arXiv:1712.05526}.

\bibitem[{Dhillon et~al.(2019)Dhillon, Chaudhari, Ravichandran, and
  Soatto}]{dhillon2019baseline}
Dhillon, G.~S.; Chaudhari, P.; Ravichandran, A.; and Soatto, S. 2019.
\newblock A baseline for few-shot image classification.
\newblock \emph{ICLR}.

\bibitem[{Doersch, Gupta, and Zisserman(2020)}]{doersch2020crosstransformers}
Doersch, C.; Gupta, A.; and Zisserman, A. 2020.
\newblock Crosstransformers: spatially-aware few-shot transfer.
\newblock \emph{NeurIPS}.

\bibitem[{Finn, Abbeel, and Levine(2017)}]{finn2017model}
Finn, C.; Abbeel, P.; and Levine, S. 2017.
\newblock Model-agnostic meta-learning for fast adaptation of deep networks.
\newblock In \emph{ICML}.

\bibitem[{Gao et~al.(2023)Gao, Bai, Gu, Yang, and Xia}]{gao2023backdoor}
Gao, K.; Bai, Y.; Gu, J.; Yang, Y.; and Xia, S.-T. 2023.
\newblock Backdoor Defense via Adaptively Splitting Poisoned Dataset.
\newblock In \emph{CVPR}.

\bibitem[{Gordon et~al.(2018)Gordon, Bronskill, Bauer, Nowozin, and
  Turner}]{gordon2018versa}
Gordon, J.; Bronskill, J.; Bauer, M.; Nowozin, S.; and Turner, R.~E. 2018.
\newblock Versa: Versatile and efficient few-shot learning.
\newblock In \emph{NeurIPS}.

\bibitem[{Gu et~al.(2018)Gu, Wang, Kuen, Ma, Shahroudy, Shuai, Liu, Wang, Wang,
  Cai et~al.}]{gu2018recent}
Gu, J.; Wang, Z.; Kuen, J.; Ma, L.; Shahroudy, A.; Shuai, B.; Liu, T.; Wang,
  X.; Wang, G.; Cai, J.; et~al. 2018.
\newblock Recent advances in convolutional neural networks.
\newblock \emph{Pattern Recognition}.

\bibitem[{Gu et~al.(2023)Gu, Wei, Torr, and Hu}]{gu2023exploring}
Gu, J.; Wei, F.; Torr, P.; and Hu, H. 2023.
\newblock Exploring Non-additive Randomness on ViT against Query-Based
  Black-Box Attacks.
\newblock \emph{BMVC}.

\bibitem[{Gu et~al.(2022)Gu, Zhao, Tresp, and Torr}]{gu2022segpgd}
Gu, J.; Zhao, H.; Tresp, V.; and Torr, P.~H. 2022.
\newblock SegPGD: An Effective and Efficient Adversarial Attack for Evaluating
  and Boosting Segmentation Robustness.
\newblock In \emph{ECCV}.

\bibitem[{Gu et~al.(2019)Gu, Liu, Dolan-Gavitt, and Garg}]{gu2019badnets}
Gu, T.; Liu, K.; Dolan-Gavitt, B.; and Garg, S. 2019.
\newblock Badnets: Evaluating backdooring attacks on deep neural networks.
\newblock \emph{IEEE Access}.

\bibitem[{Guan et~al.(2022)Guan, Tu, He, and Tao}]{guan2022few}
Guan, J.; Tu, Z.; He, R.; and Tao, D. 2022.
\newblock Few-shot backdoor defense using shapley estimation.
\newblock In \emph{CVPR}.

\bibitem[{Gudibande et~al.()Gudibande, Chen, Bai, Xiong, and
  Song}]{li2022untargeted}
Gudibande, A.; Chen, X.; Bai, Y.; Xiong, J.; and Song, D. ????
\newblock Test-time Adaptation of Residual Blocks against Poisoning and
  Backdoor Attacks.

\bibitem[{He et~al.(2023)He, Liu, Li, Liang, Li, Jia, and
  Cao}]{he2023generating}
He, B.; Liu, J.; Li, Y.; Liang, S.; Li, J.; Jia, X.; and Cao, X. 2023.
\newblock Generating transferable 3d adversarial point cloud via random
  perturbation factorization.
\newblock In \emph{AAAI}.

\bibitem[{He et~al.(2016)He, Zhang, Ren, and Sun}]{he2016deep}
He, K.; Zhang, X.; Ren, S.; and Sun, J. 2016.
\newblock Deep residual learning for image recognition.

\bibitem[{Huang et~al.(2021)Huang, Guo, Juefei-Xu, Ma, Miao, Liu, and
  Pu}]{huang2021advfilter}
Huang, Y.; Guo, Q.; Juefei-Xu, F.; Ma, L.; Miao, W.; Liu, Y.; and Pu, G. 2021.
\newblock AdvFilter: predictive perturbation-aware filtering against
  adversarial attack via multi-domain learning.
\newblock In \emph{ACM MM}, 395--403.

\bibitem[{Huang et~al.(2023)Huang, Sun, Guo, Juefei-Xu, Zhu, Feng, Liu, and
  Pu}]{huang2023ala}
Huang, Y.; Sun, L.; Guo, Q.; Juefei-Xu, F.; Zhu, J.; Feng, J.; Liu, Y.; and Pu,
  G. 2023.
\newblock ALA: Naturalness-aware Adversarial Lightness Attack.
\newblock In \emph{ACM MM}, 2418--2426.

\bibitem[{Jia et~al.(2020)Jia, Wei, Cao, and Han}]{jia2020adv}
Jia, X.; Wei, X.; Cao, X.; and Han, X. 2020.
\newblock Adv-watermark: A novel watermark perturbation for adversarial
  examples.
\newblock In \emph{ACM MM}, 1579--1587.

\bibitem[{Kang et~al.(2021)Kang, Kwon, Min, and Cho}]{kang2021relational}
Kang, D.; Kwon, H.; Min, J.; and Cho, M. 2021.
\newblock Relational Embedding for Few-Shot Classification.
\newblock In \emph{ICCV}.

\bibitem[{Koch et~al.(2015)Koch, Zemel, Salakhutdinov et~al.}]{koch2015siamese}
Koch, G.; Zemel, R.; Salakhutdinov, R.; et~al. 2015.
\newblock Siamese neural networks for one-shot image recognition.
\newblock In \emph{ICML deep learning workshop}.

\bibitem[{Lee et~al.(2019)Lee, Maji, Ravichandran, and Soatto}]{lee2019meta}
Lee, K.; Maji, S.; Ravichandran, A.; and Soatto, S. 2019.
\newblock Meta-learning with differentiable convex optimization.
\newblock In \emph{CVPR}.

\bibitem[{Li et~al.(2021)Li, Dong, Tian, Qin, Yang, Wang, Huo, Shi, Wang, Gao
  et~al.}]{li2021libfewshot}
Li, W.; Dong, C.; Tian, P.; Qin, T.; Yang, X.; Wang, Z.; Huo, J.; Shi, Y.;
  Wang, L.; Gao, Y.; et~al. 2021.
\newblock LibFewShot: A Comprehensive Library for Few-shot Learning.
\newblock \emph{arXiv preprint arXiv:2109.04898}.

\bibitem[{Li et~al.(2019{\natexlab{a}})Li, Wang, Xu, Huo, Gao, and
  Luo}]{li2019revisiting}
Li, W.; Wang, L.; Xu, J.; Huo, J.; Gao, Y.; and Luo, J. 2019{\natexlab{a}}.
\newblock Revisiting local descriptor based image-to-class measure for few-shot
  learning.
\newblock In \emph{CVPR}.

\bibitem[{Li et~al.(2019{\natexlab{b}})Li, Wang, Zhang, Huo, Gao, and
  Luo}]{li2019defensive}
Li, W.; Wang, L.; Zhang, X.; Huo, J.; Gao, Y.; and Luo, J. 2019{\natexlab{b}}.
\newblock Defensive Few-shot Adversarial Learning.
\newblock \emph{arXiv preprint arXiv:1911.06968}.

\bibitem[{Li et~al.(2023)Li, Ya, Bai, Jiang, and Xia}]{li2023backdoorbox}
Li, Y.; Ya, M.; Bai, Y.; Jiang, Y.; and Xia, S.-T. 2023.
\newblock BackdoorBox: A python toolbox for backdoor learning.
\newblock \emph{arXiv preprint arXiv:2302.01762}.

\bibitem[{Li et~al.(2022{\natexlab{a}})Li, Zhong, Ma, Jiang, and
  Xia}]{li2022few}
Li, Y.; Zhong, H.; Ma, X.; Jiang, Y.; and Xia, S.-T. 2022{\natexlab{a}}.
\newblock Few-shot backdoor attacks on visual object tracking.
\newblock \emph{ICLR}.

\bibitem[{Li et~al.(2022{\natexlab{b}})Li, Zhu, Jia, Jiang, Xia, and
  Cao}]{li2022defending}
Li, Y.; Zhu, L.; Jia, X.; Jiang, Y.; Xia, S.-T.; and Cao, X.
  2022{\natexlab{b}}.
\newblock Defending against model stealing via verifying embedded external
  features.
\newblock In \emph{AAAI}.

\bibitem[{Liang et~al.(2022)Liang, Li, Fan, Jia, Li, Wu, and
  Cao}]{liang2022large}
Liang, S.; Li, L.; Fan, Y.; Jia, X.; Li, J.; Wu, B.; and Cao, X. 2022.
\newblock A large-scale multiple-objective method for black-box attack against
  object detection.
\newblock In \emph{ECCV}.

\bibitem[{Liu et~al.(2020{\natexlab{a}})Liu, Cao, Lin, Li, Zhang, Long, and
  Hu}]{liu2020negative}
Liu, B.; Cao, Y.; Lin, Y.; Li, Q.; Zhang, Z.; Long, M.; and Hu, H.
  2020{\natexlab{a}}.
\newblock Negative margin matters: Understanding margin in few-shot
  classification.
\newblock In \emph{ECCV}.

\bibitem[{Liu, Dolan-Gavitt, and Garg(2018)}]{liu2018fine}
Liu, K.; Dolan-Gavitt, B.; and Garg, S. 2018.
\newblock Fine-pruning: Defending against backdooring attacks on deep neural
  networks.
\newblock In \emph{Proc. of RAID}.

\bibitem[{Liu et~al.(2022)Liu, Liu, Bai, Gu, Chen, Jia, and
  Cao}]{liu2022watermark}
Liu, X.; Liu, J.; Bai, Y.; Gu, J.; Chen, T.; Jia, X.; and Cao, X. 2022.
\newblock Watermark Vaccine: Adversarial Attacks to Prevent Watermark Removal.
\newblock In \emph{ECCV}.

\bibitem[{Liu et~al.(2020{\natexlab{b}})Liu, Ma, Bailey, and
  Lu}]{liu2020reflection}
Liu, Y.; Ma, X.; Bailey, J.; and Lu, F. 2020{\natexlab{b}}.
\newblock Reflection backdoor: A natural backdoor attack on deep neural
  networks.
\newblock In \emph{ECCV}.

\bibitem[{Liu, Xie, and Srivastava(2017)}]{liu2017neural}
Liu, Y.; Xie, Y.; and Srivastava, A. 2017.
\newblock Neural trojans.
\newblock In \emph{ICCD}.

\bibitem[{Madry et~al.(2018)Madry, Makelov, Schmidt, Tsipras, and
  Vladu}]{madry2017towards}
Madry, A.; Makelov, A.; Schmidt, L.; Tsipras, D.; and Vladu, A. 2018.
\newblock Towards deep learning models resistant to adversarial attacks.
\newblock In \emph{ICLR Poster}.

\bibitem[{Nguyen and Tran(2021)}]{nguyen2021wanet}
Nguyen, A.; and Tran, A. 2021.
\newblock WaNet--Imperceptible Warping-based Backdoor Attack.
\newblock \emph{ICLR}.

\bibitem[{Nguyen and Tran(2020)}]{nguyen2020input}
Nguyen, T.~A.; and Tran, A. 2020.
\newblock Input-aware dynamic backdoor attack.
\newblock \emph{NeurIPS}.

\bibitem[{Radford et~al.(2021)Radford, Kim, Hallacy, Ramesh, Goh, Agarwal,
  Sastry, Askell, Mishkin, Clark et~al.}]{radford2021learning}
Radford, A.; Kim, J.~W.; Hallacy, C.; Ramesh, A.; Goh, G.; Agarwal, S.; Sastry,
  G.; Askell, A.; Mishkin, P.; Clark, J.; et~al. 2021.
\newblock Learning transferable visual models from natural language
  supervision.
\newblock In \emph{ICML}.

\bibitem[{Raghu et~al.(2020)Raghu, Raghu, Bengio, and Vinyals}]{raghu2020rapid}
Raghu, A.; Raghu, M.; Bengio, S.; and Vinyals, O. 2020.
\newblock Rapid learning or feature reuse? towards understanding the
  effectiveness of maml.
\newblock \emph{ICLR}.

\bibitem[{Rajasegaran et~al.(2020)Rajasegaran, Khan, Hayat, Khan, and
  Shah}]{rajasegaran2020self}
Rajasegaran, J.; Khan, S.; Hayat, M.; Khan, F.~S.; and Shah, M. 2020.
\newblock Self-supervised knowledge distillation for few-shot learning.
\newblock \emph{arXiv preprint arXiv:2006.09785}.

\bibitem[{Ren et~al.(2018)Ren, Triantafillou, Ravi, Snell, Swersky, Tenenbaum,
  Larochelle, and Zemel}]{ren2018meta}
Ren, M.; Triantafillou, E.; Ravi, S.; Snell, J.; Swersky, K.; Tenenbaum, J.~B.;
  Larochelle, H.; and Zemel, R.~S. 2018.
\newblock Meta-learning for semi-supervised few-shot classification.
\newblock \emph{ICLR}.

\bibitem[{Rusu et~al.(2019{\natexlab{a}})Rusu, Rao, Sygnowski, Vinyals,
  Pascanu, Osindero, and Hadsell}]{rusu2019meta}
Rusu, A.~A.; Rao, D.; Sygnowski, J.; Vinyals, O.; Pascanu, R.; Osindero, S.;
  and Hadsell, R. 2019{\natexlab{a}}.
\newblock Meta-learning with latent embedding optimization.
\newblock \emph{ICLR}.

\bibitem[{Rusu et~al.(2019{\natexlab{b}})Rusu, Rao, Sygnowski, Vinyals,
  Pascanu, Osindero, and Hadsell}]{rusu2018meta}
Rusu, A.~A.; Rao, D.; Sygnowski, J.; Vinyals, O.; Pascanu, R.; Osindero, S.;
  and Hadsell, R. 2019{\natexlab{b}}.
\newblock Meta-learning with latent embedding optimization.
\newblock \emph{ICLR}.

\bibitem[{Saha, Subramanya, and Pirsiavash(2020)}]{saha2020hidden}
Saha, A.; Subramanya, A.; and Pirsiavash, H. 2020.
\newblock Hidden trigger backdoor attacks.
\newblock In \emph{AAAI}.

\bibitem[{Snell, Swersky, and Zemel(2017)}]{snell2017prototypical}
Snell, J.; Swersky, K.; and Zemel, R. 2017.
\newblock Prototypical networks for few-shot learning.
\newblock \emph{NeurIPS}.

\bibitem[{Sung et~al.(2018)Sung, Yang, Zhang, Xiang, Torr, and
  Hospedales}]{sung2018learning}
Sung, F.; Yang, Y.; Zhang, L.; Xiang, T.; Torr, P.~H.; and Hospedales, T.~M.
  2018.
\newblock Learning to compare: Relation network for few-shot learning.
\newblock In \emph{CVPR}.

\bibitem[{Tian et~al.(2020)Tian, Wang, Krishnan, Tenenbaum, and
  Isola}]{tian2020rethinking}
Tian, Y.; Wang, Y.; Krishnan, D.; Tenenbaum, J.~B.; and Isola, P. 2020.
\newblock Rethinking few-shot image classification: a good embedding is all you
  need?
\newblock In \emph{ECCV}.

\bibitem[{Turner, Tsipras, and Madry(2019)}]{turner2019label}
Turner, A.; Tsipras, D.; and Madry, A. 2019.
\newblock Label-consistent backdoor attacks.
\newblock \emph{arXiv preprint arXiv:1912.02771}.

\bibitem[{Vinyals et~al.(2016)Vinyals, Blundell, Lillicrap, Wierstra
  et~al.}]{vinyals2016matching}
Vinyals, O.; Blundell, C.; Lillicrap, T.; Wierstra, D.; et~al. 2016.
\newblock Matching networks for one shot learning.
\newblock \emph{NeuraIPS}.

\bibitem[{Wang et~al.(2019)Wang, Yao, Shan, Li, Viswanath, Zheng, and
  Zhao}]{wang2019neural}
Wang, B.; Yao, Y.; Shan, S.; Li, H.; Viswanath, B.; Zheng, H.; and Zhao, B.~Y.
  2019.
\newblock Neural cleanse: Identifying and mitigating backdoor attacks in neural
  networks.
\newblock In \emph{Proc. of IEEE S$\&$P}. IEEE.

\bibitem[{Weiss, Khoshgoftaar, and Wang(2016)}]{weiss2016survey}
Weiss, K.; Khoshgoftaar, T.~M.; and Wang, D. 2016.
\newblock A survey of transfer learning.
\newblock \emph{Journal of Big data}.

\bibitem[{Wertheimer, Tang, and Hariharan(2021)}]{wertheimer2021few}
Wertheimer, D.; Tang, L.; and Hariharan, B. 2021.
\newblock Few-shot classification with feature map reconstruction networks.
\newblock In \emph{CVPR}.

\bibitem[{Xu et~al.(2020)Xu, Wang, Tu et~al.}]{xu2020attentional}
Xu, W.; Wang, H.; Tu, Z.; et~al. 2020.
\newblock Attentional constellation nets for few-shot learning.
\newblock In \emph{ICLR}.

\bibitem[{Yang, Liu, and Xu(2021)}]{yang2021free}
Yang, S.; Liu, L.; and Xu, M. 2021.
\newblock Free lunch for few-shot learning: Distribution calibration.
\newblock \emph{ICLR}.

\bibitem[{Zhang et~al.(2020)Zhang, Cai, Lin, and Shen}]{zhang2020deepemd}
Zhang, C.; Cai, Y.; Lin, G.; and Shen, C. 2020.
\newblock Deepemd: Few-shot image classification with differentiable earth
  mover's distance and structured classifiers.
\newblock In \emph{CVPR}.

\bibitem[{Zhao et~al.(2020)Zhao, Ma, Zheng, Bailey, Chen, and
  Jiang}]{zhao2020clean}
Zhao, S.; Ma, X.; Zheng, X.; Bailey, J.; Chen, J.; and Jiang, Y.-G. 2020.
\newblock Clean-label backdoor attacks on video recognition models.
\newblock In \emph{CVPR}.

\end{thebibliography}

\newpage
\section{Appendix}


\subsection{Description of Datasets}
\label{sec:data}
In this paper, we evaluate our backdoor attack on two widely-used benchmark datasets: \textit{mini}ImageNet and \textit{tiered}ImageNet. Table~\ref{dataset} shows the data split for each dataset. The \textit{mini}ImageNet~\cite{vinyals2016matching} is a subset of ImageNet that contains 100 categories, with 600 images for each class. The \textit{tiered}ImageNet~\cite{ren2018meta} is also a subset of ImageNet, with 608 classes and 776,165 images. In LibFewshot, these datasets can be split into a training set, a validation set, and a testing set, and we follow the data split setting in LibFewshot \cite{li2022few}. We present the attack results of \textit{mini}ImageNet in the manuscript, while the results of \textit{tiered}ImageNet~\cite{ren2018meta} are shown in Sec.~\ref{sec:tiered} of this supplementary material.

\subsection{Detailed Settings for Training}
In the main experiments of the manuscript, we evaluate our FLBA on three typical methods of few-shot learning: Baseline++~\cite{chen2019closer}, MAML~\cite{finn2017model}, and  ProtoNet~\cite{snell2017prototypical}. Here we will give their detailed settings at the training time. 
In the following sections, we provide a detailed description of their settings during the training phase. All experiments are conducted based on the open-source code available.\footnote{https://github.com/RL-VIG/LibFewShot}

\subsection{Setting for Baseline++} 
We adopt the Baseline++ model provided in LibFewShot and use the Stochastic Gradient Descent (SGD) optimizer with an initial learning rate of 0.01, momentum of 0.9, and weight decay of 0.0005 to train on the auxiliary set. We train the model for 100 epochs with a ResNet12 backbone on a single NVIDIA 2080Ti GPU. During training, we set the way number to 5, shot number to 5, and the query number to 15 for a 5-shot learning. We use the SGD optimizer with a learning rate of 0.01, momentum of 0.9, and weight decay of 0.001 for 100 iterations to train the classifier. Data augmentation is applied once during the training and testing stages.

\subsection{Setting for MAML}
 For the MAML model, we utilized the Adam optimizer with an initial learning rate of 0.01 to train on the auxiliary set. We trained the model for 100 epochs with a ResNet12 backbone on a single NVIDIA 2080Ti. During the training, we chose the way number, shot number, and query number all as 5 for a 5-shot learning and set the episode as 2. We used the Adam optimizer with a learning rate of 0.01 for the second phase training. Data augmentation was employed during both the training and testing stages. The train episode was set to 2000, and the test episode was set to 600.

\subsection{Setting for ProtoNet}
For the ProtoNet model, we followed the settings in Libfewshot and used the Adam optimizer with an initial learning rate of 0.001 and a weight decay of 0.00005 to train it. We trained the model for 100 epochs with a ResNet12 backbone on a single NVIDIA 2080Ti. During the training, we chose the way number, shot number, and query number as 5 for a 5-shot learning. Data augmentation was applied during both the training and testing stages. The train episode was set to 2000, and the test episode was set to 600.

\begin{table}[t]
\begin{center}
\resizebox{0.8\linewidth}{!}{
\begin{tabular}{@{}c|cc|cc@{}}
\toprule
Dataset & \multicolumn{2}{c|}{miniImageNet} & \multicolumn{2}{c}{tieredImageNet}  \\
Num & $N_{class}$ & $N_{images}$ & $N_{class}$ & $N_{images}$  \\ \midrule
Train & 64 & 38400 & 351 & 448695 \\
Val & 16 & 9600 & 97 & 124261  \\
Test & 20 & 12000 & 160 & 206209  \\ \midrule
All & 100 & 60000 & 608 & 779165  \\ \bottomrule
\end{tabular}}
\end{center}
\caption{The data split on two benchmarks datasets.$N_{class}$ is the number of classes in training (auxiliary), validation and test sets. $N_{images}$  is the number of images in them. And the bottom row in the table shows the total number of categories and images.}
\label{dataset}
\end{table}

\begin{table*}[t]
\begin{center}
\tabcolsep=0.6cm
\resizebox{0.8\linewidth}{!}{
\begin{tabular}{cc|cc|cc}
\toprule
\multicolumn{2}{c|}{\multirow{2}{*}{Methods}}   & \multicolumn{2}{c|}{Clean} & \multicolumn{2}{c}{FLBA} \\
\multicolumn{2}{c|}{}                           & ASR          & BA          & ASR           & BA           \\ \midrule
\multirow{2}{*}{Fine-tuning based methods}     & SKD~\cite{rajasegaran2020self}    & 20.2         & 66.4        & 60.4          & 60.2         \\
                                       & Negcos~\cite{liu2020negative} & 19.9         & 78.7        & 82.0          & \textbf{67.4}         \\ \midrule
\multirow{2}{*}{Meta-learning based methods}   & R2D2~\cite{bertinetto2018meta}   & 19.7         & 73.7        & \textbf{87.3}          & 61.1         \\
                                       & LEO~\cite{rusu2018meta}    & 19.9         & 69.3        & 79.3          & 53.9         \\
                                       \midrule
\multirow{2}{*}{Metric-learning based methods} & DN4~\cite{li2019revisiting}    & 20.0         & 75.2        & 60.6          & 62.3         \\
                                       & RENet~\cite{kang2021relational}  & 21.1         & 69.1        & 68.2          & 66.7         \\ \bottomrule
\end{tabular}}
\end{center}
\caption{Attack results of our FLBA on \textit{mini}ImageNet across three types of few-shot learning paradigms with six popular methods. In each case of FLBA, the best ASR and BA are \textbf{boldfaced.}}
\label{tab:more models}
\end{table*}

\subsection{Effectiveness on More Models}

In our manuscript, we demonstrate the successful construction of backdoors in three types of few-shot learning networks using our FLBA. To further highlight the effectiveness of our attacks, we also evaluate six additional few-shot learning methods that belong to latest different paradigms. These include SKD~\cite{rajasegaran2020self} and Negcos~\cite{liu2020negative} for fine-tuning based approaches, R2D2~\cite{bertinetto2018meta} and LEO~\cite{rusu2018meta} for meta-learning based methods, and DN4~\cite{li2019revisiting} and RENet~\cite{kang2021relational} for metric-learning based approaches.

As shown in Table~\ref{tab:more models}, our FLBA successfully applies to all the few-shot learning paradigms, and the attack success rates exceed 60$\%$ while maintaining a good BA compared to their clean results. Notably, R2D2 exhibits the highest vulnerability to our backdoor attack, with an attack success rate of 87.3$\%$, while SKD displays some resilience to our attack, with an ASR of only 60.4$\%$. In conclusion, our backdoor attack demonstrates a high level of universality across diverse few-shot learning methods.

\subsection{Effectiveness on CLIP}
Recently, multimodal contrastive learning methods such as CLIP~\cite{radford2021learning} have shown remarkable results as pre-trained encoders for zero-shot and few-shot learning. In this section, we explore the effectiveness of our FLBA on several pre-trained image encoders of CLIP. We test four different backbones, including ResNet50, ResNet101, ViT-B/32, and ViT-B/16, and the attack results are shown in Table~\ref{tab:clip}. It is surprising to note that our FLBA achieves significantly higher ASR and BA on CLIP encoders compared to the embedding pre-trained on miniImageNet, as reported in the manuscript. Specifically, the ASR for ViT-B/32 can almost reach 100$\%$, while maintaining a high BA of 95.4. This observation can be attributed to the fact that CLIP obtains an excellent encoder on a larger dataset, which shows a good distribution of features of different class. Thus, the optimized triggers of FLBA based on this encoder can effectively move away from the normal distribution, leading to an effective backdoor attack.

\begin{table}[t]
\begin{center}
\tabcolsep=0.6cm
\resizebox{0.7\linewidth}{!}{
\begin{tabular}{c|cc}
\toprule
Backbone & ASR  & BA   \\ \midrule
RN50     & 94.4 & 86.6 \\
RN101    & 81.8 & 91.1 \\
ViT-B/32 & \textbf{99.9} & 95.4 \\
ViT-B/16 & 99.8 & \textbf{95.9} \\ \bottomrule
\end{tabular}}
\end{center}
\caption{Results of our FLBA on \textit{mini}ImageNet with four image encoder backbones of CLIP. In each case, the best ASR and BA are \textbf{boldfaced.}}
\label{tab:clip}
\end{table}

\subsection{Effectiveness on the \textit{tiered}ImageNet}
\label{sec:tiered}
In the manuscript, we have conducted a series of experiments utilizing the \textit{mini}ImageNet dataset. To ensure the robustness and generalizability of our approach, we have evaluated its effect on a different few-shot learning dataset, namely, \textit{tiered}ImageNet~\cite{ren2018meta}, which has been introduced in Sec.~\ref{sec:data} of the Supplementary Material. As depicted in Table~\ref{tab:tiered}, prior backdoor attack methods exhibited inferior performance on the \textit{tiered}ImageNet dataset compared to our proposed method, the FLBA, which achieved an impressive Attack Success Rate (ASR) of 99.2. This is consistent with the performance on \textit{mini}ImageNet in the manuscript. However, it is noteworthy that our FLBA suffered a little decrease in Backdoor Accuracy (BA) compared to the results on \textit{mini}ImageNet. Despite it, our approach still outperforms existing backdoor attack methods and exhibits dataset-independent effectiveness.

\begin{table}[t]
\begin{center}
\tabcolsep=0.6cm
\resizebox{0.8\linewidth}{!}{
\begin{tabular}{c|cc}
\toprule
{Method} & ASR          & BA \\ \midrule 
Clean                   & 19.5           & 75.1          \\ \midrule
BadNet                  & 55.9           & 56.9          \\
Blended                 & 65.0           & 56.6          \\
Refool                  & 75.6           & 64.8          \\
WaNet                   & 71.2           & 48.7          \\
LabelConsistent         & 15.8           & 42.7          \\
TUAP                    & 21.2           & 64.7          \\
\textbf{Ours}           & \textbf{99.2}  & \textbf{66.3} \\ \bottomrule
\end{tabular}}
\end{center}
\caption{Comparison($\%$) of different backdoor attack methods on \textit{tiered}ImageNet. In each case, the best attacking ASR and BA are \textbf{boldfaced.}}
\label{tab:tiered}
\end{table}

\subsection{Transferability}
\begin{figure}[t]
    \centering
    \includegraphics[width=0.9\linewidth]{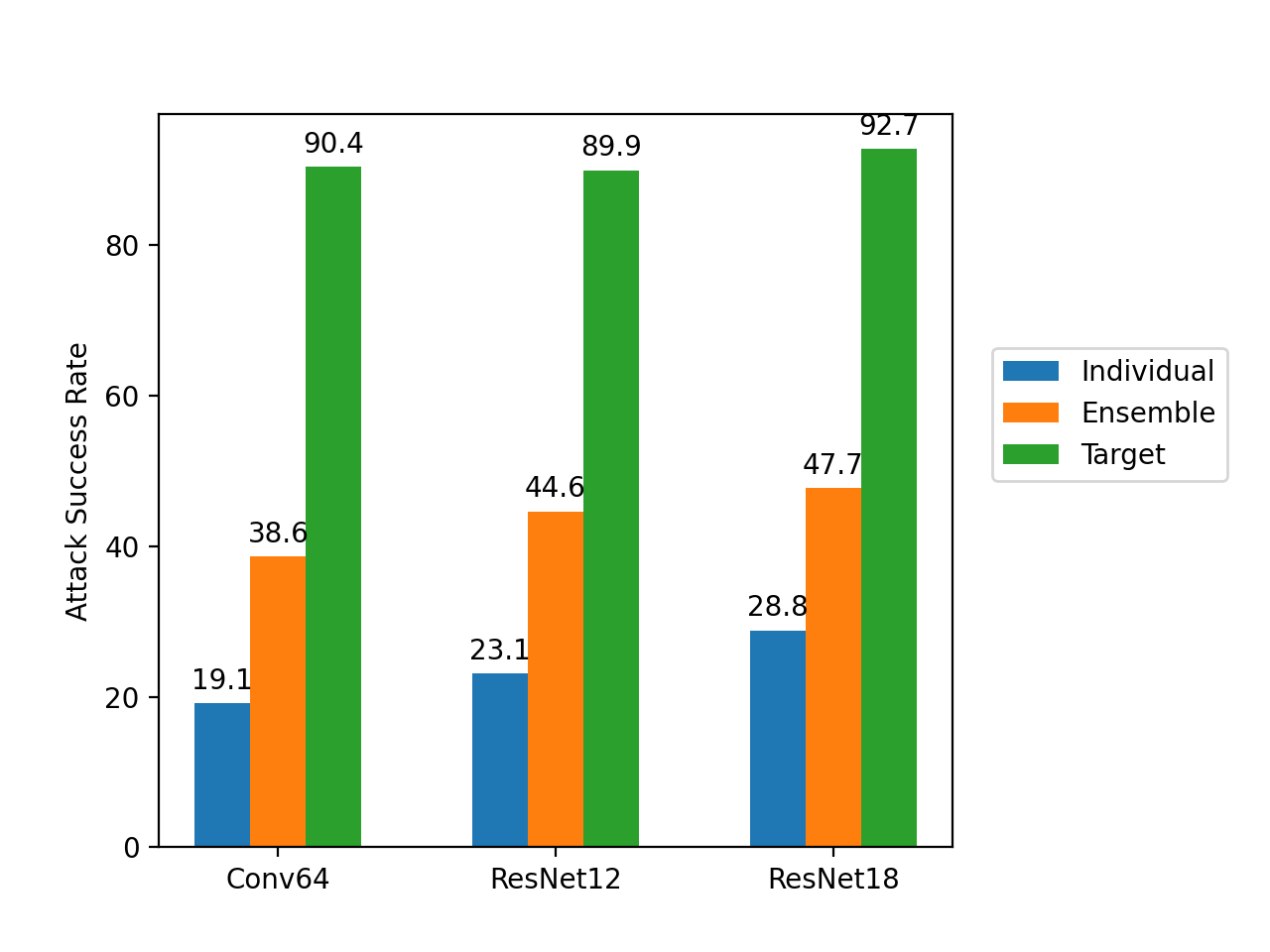}
    \caption{The transferability results of our FLBA. An individual poisoned dataset generated by another source embedding can not build a backdoor successfully on target embeddings, while the ensemble poisoned dataset generated by several embeddings can improve the ASR significantly. }
    \label{fig:trans}
\end{figure}

The present study assumes that the attacker has access to the pre-trained model of the victim in the training support set. However, in some scenarios, the attacker may lack information about the victim and cannot obtain the same pre-trained embedding. In this section, we investigate the transferability of different pre-trained embeddings. Specifically, we examine three pre-trained embeddings: Conv64F, ResNet12, and ResNet18. We test the ASR on the target embedding with the poisoned dataset generated by other source embeddings. Since the poisoned dataset is optimized from only one source embedding, we call it 'Individual', and their results are illustrated in the 'Individual' bar of Fig.~\ref{fig:trans}. Our findings indicate that the poisoned dataset from the source embedding does not transfer well to other embeddings, resulting in an average success rate (ASR) of approximately 20$\%$.

Recent research has proposed ensemble perturbation to enhance the transferability in black-box scenarios~\cite{liu2022watermark}. Therefore, we explore the feasibility of merging perturbations from several different source embeddings. We observe that the ensemble poisoned dataset performs better transferability than the individual poisoned dataset, with an ASR of over 40. However, the limited transferability across different embeddings remains a significant challenge for our approach. Therefore, it is worthwhile to investigate strategies to improve the transferability of our approach.

\begin{table}[]
\begin{center}
\centering
\tabcolsep=0.6cm
\resizebox{0.8\linewidth}{!}{
\begin{tabular}{c|cc}
\toprule
\multicolumn{1}{l|}{Poisoning Type} & ASR  & BA   \\ \midrule
Target Only           & 54.0 & 67.1 \\
Untargeted Only         & 73.8 & \textbf{69.4} \\
Target +   Untargeted   & \textbf{89.1} & 65.5 \\ \bottomrule
\end{tabular}}
\end{center}
\caption{Ablation study for three types of poisoning support set.}
\label{table:balance para}
\end{table}

\begin{figure}
\begin{minipage}{0.49\linewidth}
\includegraphics[width=\linewidth]{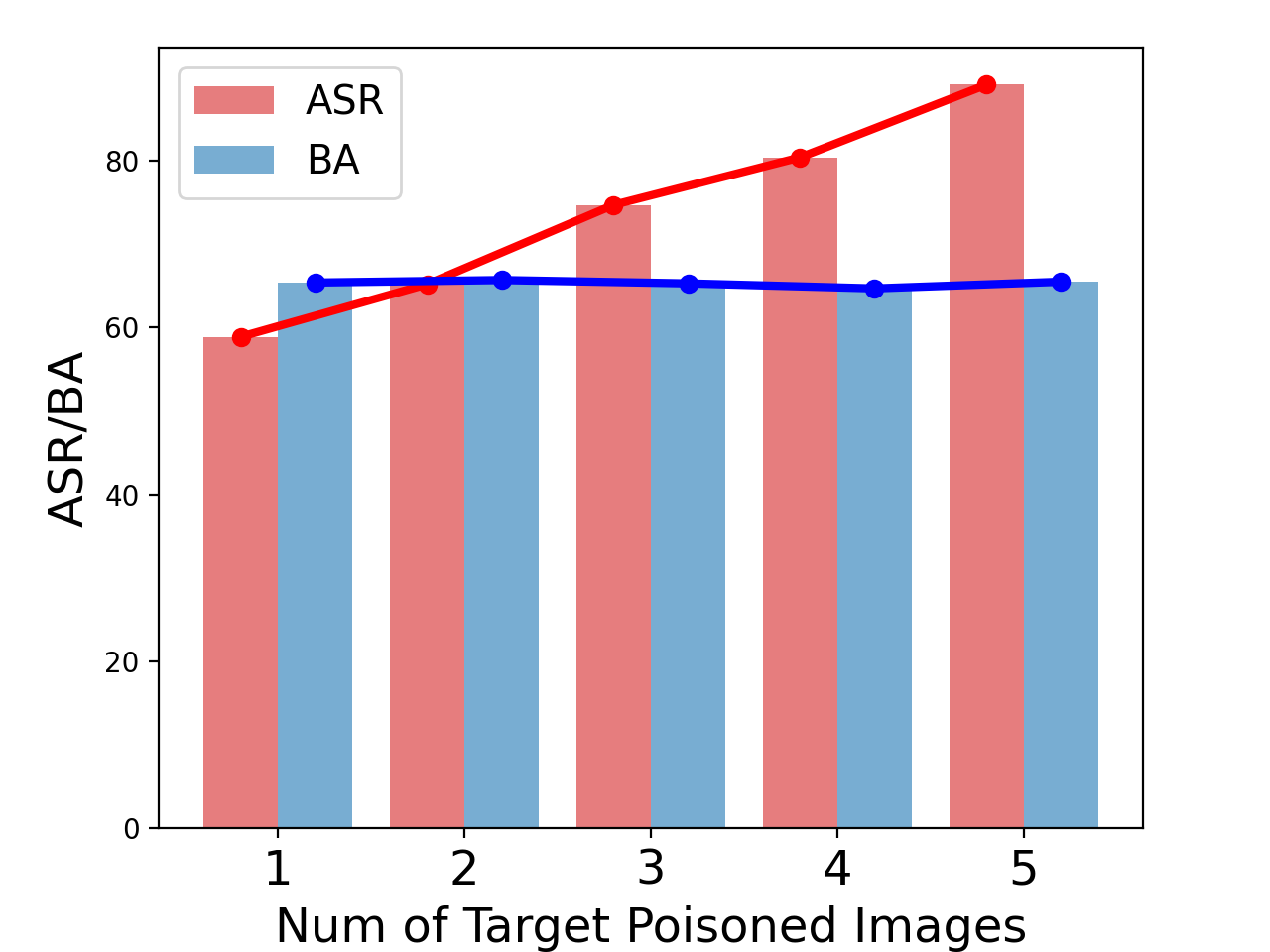}
\centerline{\small{(a) Target Class}}
\centerline{}
\end{minipage}
\hfill
\begin{minipage}{0.49\linewidth}
\includegraphics[width=\linewidth]{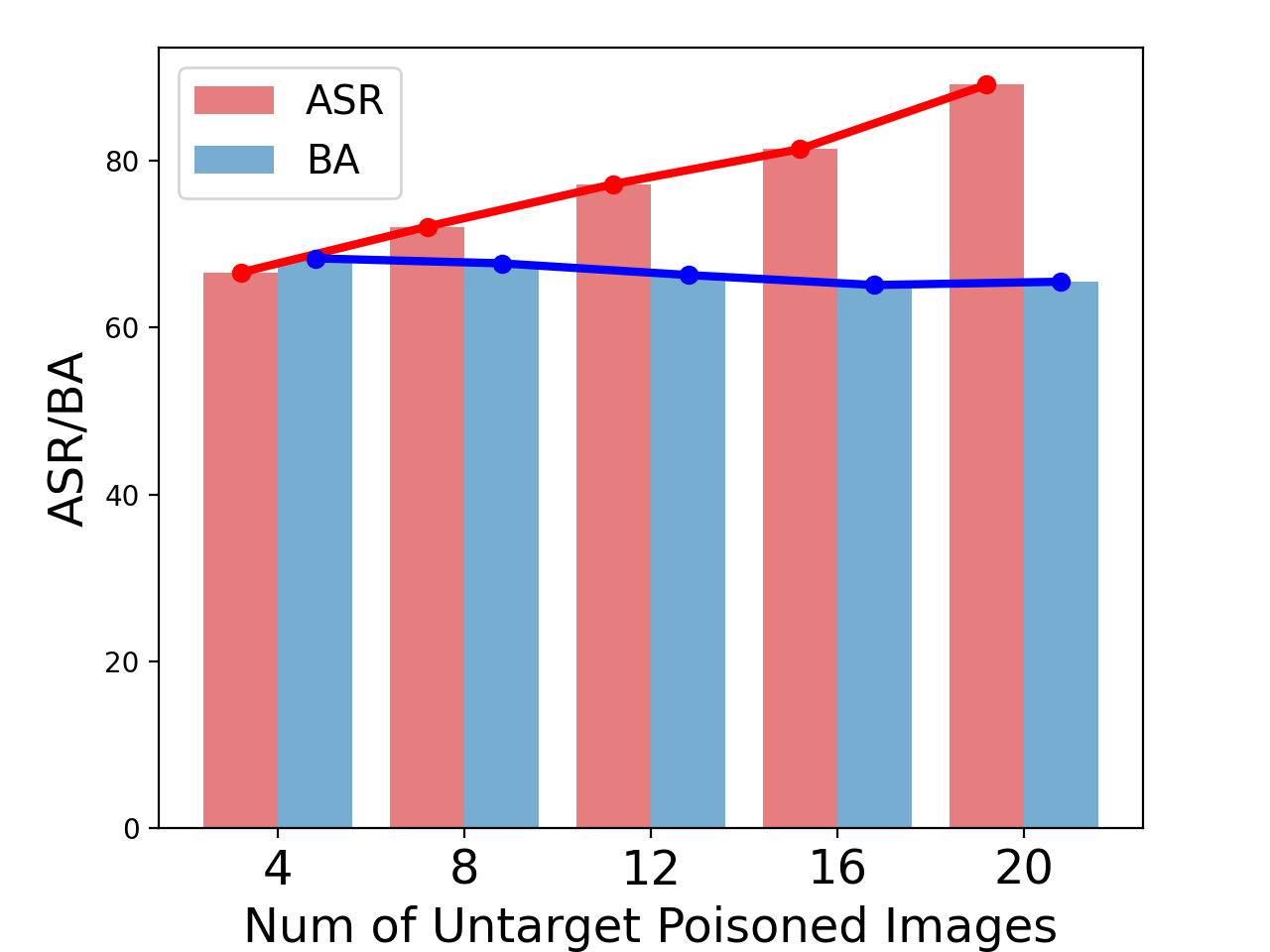}
\centerline{\small{(b) untargeted Class}}
\centerline{}
\end{minipage}
\caption{Attack results with different amount of poisoning data.}
\label{fig:ablation}
\end{figure}

\subsection{Ablation Study}
In this section, we present additional ablation studies aimed at examining the impact of poisoning size on the performance of our attack. In contrast to the primary results, which involved poisoning the entire support set, we partitioned poisoning into three categories: poisoning the target class, poisoning the untargeted class, and poisoning all classes. As presented in Table~\ref{table:balance para}, it is evident that poisoning all classes yielded the highest ASR, while poisoning the untargeted class resulted in the best BA. Consequently, to achieve better attack results, we generally opted to poison the entire dataset as a compromise. 

Furthermore, we investigated different amounts of poisoned data in the target and untargeted class, and the results are displayed in Fig~\ref{fig:ablation}. The findings revealed that as the number of poisoned images increased, the ASR for both target and untargeted classes also increased, while the BA remained almost unchanged. Therefore, it appears to be a judicious choice for our attack to forfeit the poisoning rate and poison the entire dataset to maximize the ASR.

\subsection{Sensitivity to Parameters}
In the training process, there are a few hyper-parameters for our FLBA.  In this section, we will investigate the sensitivity of FLBA to two hyper-parameters, namely trigger patterns and the size of triggers.  Other parameters are the same as those in
the main experiments of the manuscript.

\subsection{Balance Parameters}
In the experiments of the manuscript, we have two balance parameters: $\lambda_1$ and $\lambda_2$, where $\lambda_1$ controls the deviation of original features for attractive perturbations, and $\lambda_2$ controls the deviation of original features for repulsive perturbations. In most experiments, we set both parameters to 1.5. In this section, we explore the sensitivity of FLBA to these balance parameters. 

Table~\ref{table:balance para} presents the results of our experiments. We observed that decreasing the balance parameters of both attraction and repulsion perturbations consistently leads to an increase in ASR, whereas ASR decreases and BA increases when these parameters gradually increase. This observation is expected because the balance parameter determines the extent to which perturbed features deviate from their original features. Larger balance parameters result in a higher degree of consideration given to the original feature distribution, which in turn leads to a higher BA. Additionally, we noticed that the change in ASR is more pronounced than that of BA. Therefore, we finally choose a not high value of BA to achieve a higher ASR.

\begin{table}[]
\begin{center}
\centering
\tabcolsep=0.3cm
\resizebox{\linewidth}{!}{
\begin{tabular}{c|cc||c|cc}
\hline
\multicolumn{1}{l|}{Fixed $\lambda_1$ = 2} & \multicolumn{1}{l}{ASR} & \multicolumn{1}{l||}{BA} & \multicolumn{1}{l|}{Fixed $\lambda_2$ = 1.5} & \multicolumn{1}{l}{ASR}  & \multicolumn{1}{l}{BA}   \\ \hline
$\lambda_2$ = 2.5                          & 82.1                    &\textbf{ 67.1}                    & $\lambda_1$ = 2.5                            &      84.8                    &              \textbf{ 65.6 }          \\
$\lambda_2$ =2.0                           & 85.9                    & 66.2                    & $\lambda_1$ =2.0                             & 89.1                     & 65.5                     \\
$\lambda_2$ =1.5                           & 89.1                    & 65.5                    & $\lambda_1$ =1.5                             & 92.7                     & 64.2                     \\
$\lambda_2$ =1.2                           & 91.2                    & 64.1                    & $\lambda_1$ =1.2                             &                  93.4        &       63.8                   \\
$\lambda_2$ =1.0                           & \textbf{92.8 }  & 63.1   & $\lambda_1$ =1.0                             & \multicolumn{1}{l}{\textbf{94.4}} & \multicolumn{1}{l}{63.1} \\ \hline
\end{tabular}}
\end{center}
\caption{The attack results with different balance parameters. In the right column, we fix $\lambda_1$ and change only $\lambda_2$, while in the left column, we fix $\lambda_2$ but test a series of different $\lambda_1$.}
\label{table:balance para}
\end{table}

\begin{figure}[t]
\begin{minipage}{0.49\linewidth}
\includegraphics[width=\linewidth]{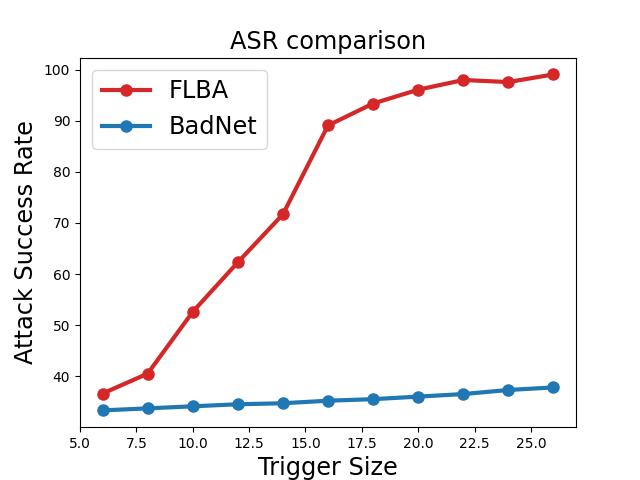}
\centerline{\small{(a) Attack Success Rate}}
\centerline{}
\end{minipage}
\hfill
\begin{minipage}{0.49\linewidth}
\includegraphics[width=\linewidth]{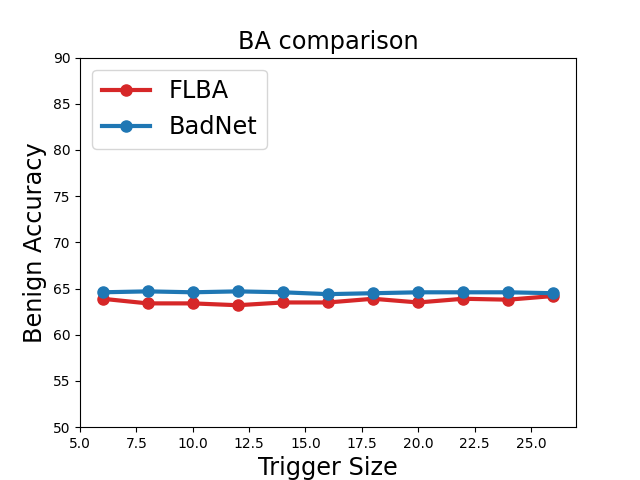}
\centerline{\small{(b) Benign Accuracy}}
\centerline{}
\end{minipage}
\caption{Attack results with different trigger size.}
\label{fig:size}
\end{figure}

\subsection{Trigger Size}
In the experiments of the manuscript, we choose the trigger size as 16 $\times$ 16. To explore the effect of trigger size on our FLBA, we conducted a series of experiments with different trigger sizes ranging from 6 $\times$ 6 to 26 $\times$ 26, while keeping the image size fixed at 84 $\times$ 84. We selected the Baseline model++ on \textit{mini}ImageNet for testing purposes, and BadNet was also used for comparison. The results are presented in Fig~\ref{fig:size}.

As we konw, a larger trigger size makes the trigger more visible to the human eye. For the attack success rate, both FLBA and BadNet's success rates increased as the trigger size increased, although FLBA's increase was more significant, ranging from 36.6$\%$ at trigger size 6 to 99.1$\%$ at trigger size 26, whereas BadNet's increase appeared to be minimal, only from 33.3$\%$ to 37.8$\%$. Thus, we believe that FLBA has greater potential for improving the attack effectiveness when the victim can tolerate a larger trigger appearing in the query image. In contrast, BadNet's attack effectiveness did not improve significantly even when it added a large trigger to the dataset. Regarding the benign accuracy, we observed that neither FLBA nor BadNet's performance was significantly affected by the trigger size. Thus, we conclude that the size of the trigger does not have a substantial impact on the BA of the infected model.

\begin{table}[]
\begin{center}
\resizebox{0.9\linewidth}{!}{
\begin{tabular}{c|cccccc}
\hline
Brightness & 0    & 0.1  & 0.2  & 0.3  & 0.4  & 0.5  \\ \hline
BA         & 64.2 & 63.9 & 63.9 & 63.7 & 63.5 & 63.4 \\
ASR        & 92.7 & 92.7 & 92.7 & 92.6 & 92.6 & 92.6 \\ \hline
\end{tabular}}
\centerline{}
\centerline{(a) Brightness}
\centerline{}
\resizebox{0.9\linewidth}{!}{
\begin{tabular}{c|cccccc}
\hline
Contrast & 0    & 0.1  & 0.2  & 0.3  & 0.4  & 0.5  \\ \hline
BA       & 64.2 & 63.9 & 64.1 & 63.9 & 64.1 & 63.8 \\
ASR      & 92.7 & 92.7 & 92.7 & 92.6 & 92.6 & 92.5 \\ \hline
\end{tabular}}
\centerline{}
\centerline{(b) Contrast}
\centerline{}

\resizebox{0.9\linewidth}{!}{
\begin{tabular}{c|cccccc}
\hline
Saturation & 0    & 0.1  & 0.2  & 0.3  & 0.4  & 0.5  \\ \hline
BA         & 64.2 & 63.8 & 63.7 & 63.5 & 63.2 & 63.1 \\
ASR        & 92.7 & 92.6 & 92.4 & 92.1 & 91.9 & 91.6 \\ \hline
\end{tabular}}
\centerline{}
\centerline{(c) Saturation}
\centerline{}

\resizebox{0.9\linewidth}{!}{
\begin{tabular}{c|cccccc}
\hline
Brightness & 0    & 0.1  & 0.2  & 0.3  & 0.4  & 0.5  \\ \hline
BA         & 64.2 & 63.9 & 63.8 & 63.6 & 63.5 & 63.4 \\
ASR        & 92.7 & 92.7 & 92.7 & 92.6 & 92.6 & 92.6 \\ \hline
\end{tabular}}
\centerline{}
\centerline{(d) Hue}
\end{center}
\caption{Resistance to four image pre-processing techniques with different budgets.}
\label{table:defense}
\end{table}

\begin{figure}[t]
\centering
    \includegraphics[width=0.8\linewidth]{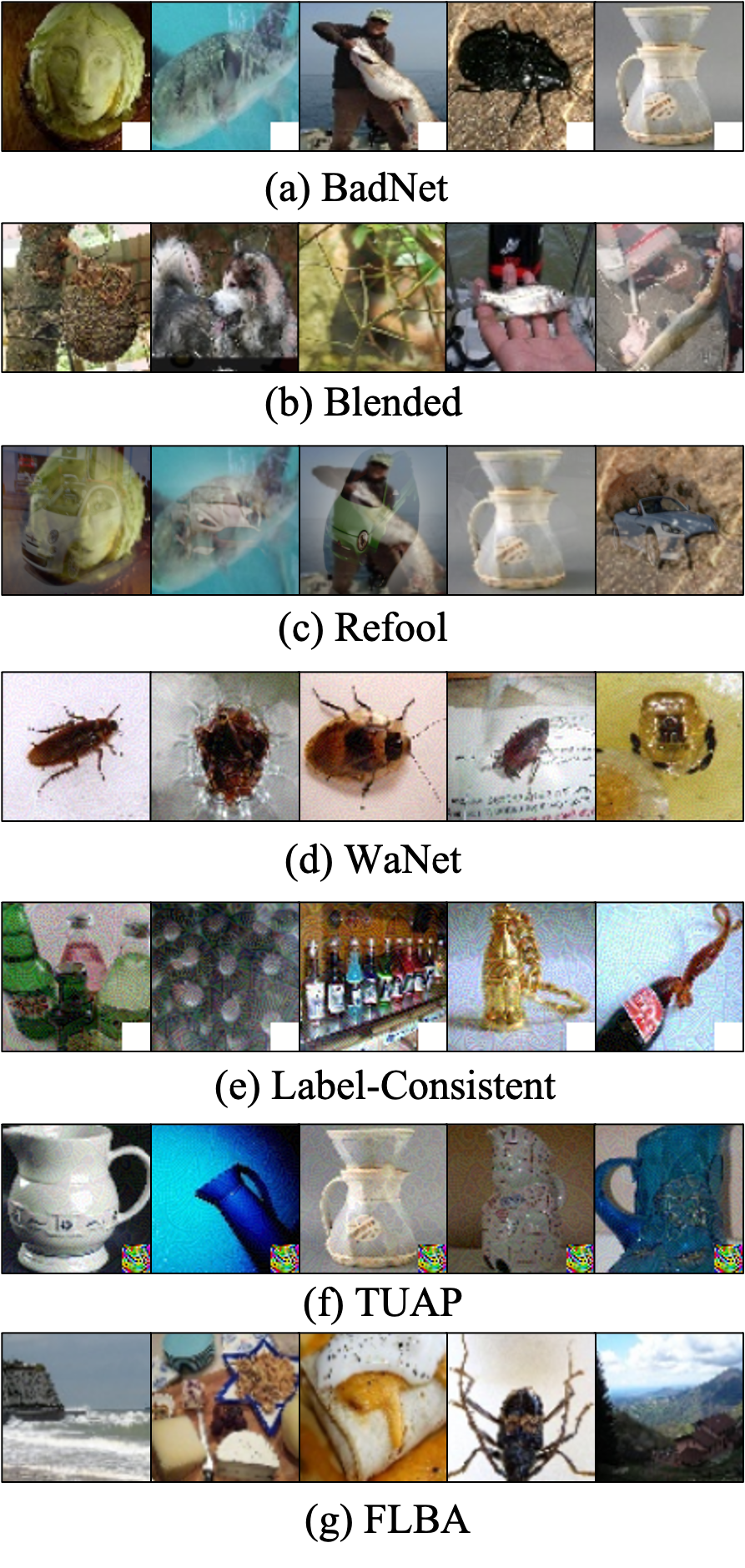}
    \caption{ The visualization of poisoned support set for different backdoor attack. }
    \label{fig:visi}
\end{figure}

\subsection{Resistance to the Image Pre-processing}
Data augmentation is a widely used technique for increasing model robustness against backdoor attacks. In this study, we employed the same four pre-processing methods used in FSBA~\cite{li2022few}, namely brightness, contrast, saturation, and hue, to evaluate the effectiveness of FLBA against defenses. The results are presented in Table~\ref{table:defense}. Our experiments showed that the four pre-processing methods had almost no effect on the backdoor attacks launched by FLBA, with the attack success rate (ASR) dropping by only about 1.0$\%$ as the adjustment budget reached 0.5. Therefore, our FLBA demonstrated successful resistance to image pre-processing defenses.

\subsection{Resistance to Neural Cleanse.} Neural Cleanse~\cite{wang2019neural} is a widely-used backdoor model mitigation method that is based on pattern optimization. The classifier is predicted to be backdoored if the Anomaly Index is larger than 2. Here, we use Neural Cleanse to detect the backdoor in an FSL classifier and adopt a clean query set in our evaluation. Table~\ref{table:NC} shows the Anomaly Index produced by Neural Cleanse for different backdoor attacks. We find that the Anomaly Indices are only 0.67 for ours, while other methods are all larger than 1 and BadNet even reaches 4.06. Thereby, our proposed attack is resistant to Neural Cleanse.

\begin{table}[]
\small
\begin{center}
\centering
\resizebox{\linewidth}{!}{
\begin{tabular}{c|ccccc}
\toprule
Backdoor Attack Method & BadNet & Blended & Refool & WaNet & Ours          \\ \midrule
Anomaly Index          & 4.64   & 3.04    & 2.25   & 1.04  & \textbf{0.67} \\ \bottomrule
\end{tabular}}
\end{center}
\caption{Anomaly Indices for the few-shot learning classifier produced by Neural Cleanse with different backdoor attacks. }
\label{table:NC}
\end{table}

\subsection{More Visualization of Different Backdoor Attacks}
In the main manuscript, we show some visualizations of the poisoned dataset regarding different types of backdoor attacks. In the supplementary material, we show more poisoned images of different backdoor attacks to better identify their stealthiness, as shown in Fig~\ref{fig:visi}. As in the manuscript, we can find that some backdoor attacks present a very obvious trigger, which is easily detectable in few-shot learning. Even though some methods such as WaNet have imperceptible triggers, they often use the dirty label method, thus its anomalies are also easily detected by the victims. However, our method (FLBA) performs better stealthiness in backdoor attacks for FSL.

\end{document}